*Review Article*

# A Tutorial on Nonorthogonal Multiple Access for 5G and Beyond

**Mahmoud Aldababsa,[1] Mesut Toka,[1,2] Selahattin Gökçeli [ID],[3] Güneş Karabulut Kurt,[3] and Oğuz Kucur [ID][1]**

[1]*Electronics Engineering Department, Gebze Technical University, Gebze, 41400 Kocaeli, Turkey*
[2]*Electrical and Electronics Engineering Department, Ömer Halisdemir University, 51240 Niğde, Turkey*
[3]*Department of Communications and Electronics Engineering, Istanbul Technical University, 34469 Istanbul, Turkey*

Correspondence should be addressed to Oğuz Kucur; okucur@gtu.edu.tr

Today's wireless networks allocate radio resources to users based on the orthogonal multiple access (OMA) principle. However, as the number of users increases, OMA based approaches may not meet the stringent emerging requirements including very high spectral efficiency, very low latency, and massive device connectivity. Nonorthogonal multiple access (NOMA) principle emerges as a solution to improve the spectral efficiency while allowing some degree of multiple access interference at receivers. In this tutorial style paper, we target providing a unified model for NOMA, including uplink and downlink transmissions, along with the extensions to multiple input multiple output and cooperative communication scenarios. Through numerical examples, we compare the performances of OMA and NOMA networks. Implementation aspects and open issues are also detailed.

## 1. Introduction

Wireless mobile communication systems became an indispensable part of modern lives. However, the number and the variety of devices increase significantly and the same radio spectrum is required to be reused several times by different applications and/or users. Additionally, the demand for the Internet of Things (IoT) introduces the necessity to connect every person and every object [1]. However, current communication systems have strict limitations, restricting any modifications and improvements on the systems to meet these demands. Recently, researchers have been working on developing suitable techniques that may be integrated in next generation wireless communication systems in order to fundamentally fulfill the emerging requirements, including very high spectral efficiency, very low latency, massive device connectivity, very high achievable data rate, ultrahigh reliability, excellent user fairness, high throughput, supporting diverse quality of services (QoS), energy efficiency, and a dramatic reduction in the cost [2]. Some potential technologies have been proposed by the academia and the industry in order to satisfy the aforementioned tight requirements and to address the challenges of future generations. For example, millimeter wave (mmWave) technology was suggested to enlarge the transmission bandwidth for very high speed communications [3], massive multiple input multiple output (MIMO) concept was presented to improve capacity and energy efficiency [4], and ultradense networks were introduced to increase the throughput and to reduce the energy consumption through using a large number of small cells [5].

Besides the aforementioned techniques, a new radio access technology is also developed by researchers to be used in communication networks due to its capability in increasing the system capacity. Recently, nonorthogonality based system designs are developed to be used in communication networks and have gained significant attention of researchers. Hence, multiple access (MA) techniques can now be fundamentally categorized as orthogonal multiple access (OMA) and nonorthogonal multiple access (NOMA). In OMA, each user can exploit orthogonal communication resources within either a specific time slot, frequency band, or code in order to avoid multiple access interference. The previous generations of networks have employed OMA schemes, such as frequency division multiple access (FDMA) of first generation (1G), time division multiple access (TDMA) of 2G, code division multiple access (CDMA) of 3G, and orthogonal frequency division multiple access (OFDMA) of 4G. In NOMA, multiple users can utilize nonorthogonal resources concurrently by yielding a high spectral efficiency while allowing some degree of multiple access interference at receivers [6, 7].

In general, NOMA schemes can be classified into two types: power-domain multiplexing and code-domain multiplexing. In power-domain multiplexing, different users are allocated different power coefficients according to their channel conditions in order to achieve a high system performance. In particular, multiple users' information signals are superimposed at the transmitter side. At the receiver side successive interference cancellation (SIC) is applied for decoding the signals one by one until the desired user's signal is obtained [8], providing a good trade-off between the throughput of the system and the user fairness. In code-domain multiplexing, different users are allocated different codes and multiplexed over the same time-frequency resources, such as multiuser shared access (MUSA) [9], sparse code multiple access (SCMA) [10], and low-density spreading (LDS) [11]. In addition to power-domain multiplexing and code-domain multiplexing, there are other NOMA schemes such as pattern division multiple access (PDMA) [12] and bit division multiplexing (BDM) [13]. Although code-domain multiplexing has a potential to enhance spectral efficiency, it requires a high transmission bandwidth and is not easily applicable to the current systems. On the other hand, power-domain multiplexing has a simple implementation as considerable changes are not required on the existing networks. Also, it does not require additional bandwidth in order to improve spectral efficiency [14]. In this review/tutorial paper, we will focus on the power-domain NOMA.

Although OMA techniques can achieve a good system performance even with simple receivers because of no mutual interference among users in an ideal setting, they still do not have the ability to address the emerging challenges due to the increasing demands in 5G networks and beyond. For example, according to International Mobile Telecommunications (IMT) for 2020 and beyond [15], 5G technology should support three main categories of scenarios, such as enhanced mobile broadband (eMBB), massive machine type communication (mMTC), and ultrareliable and low-latency communication (URLLC). The main challenging requirements of eMBB scenario are 100 Mbps user perceived data rate and more than 3 times spectrum efficiency improvement over the former LTE releases to provide services including high definition video experience, virtual reality, and augmented reality. Since a large number of IoT devices will have access to the network, the main challenge of mMTC is to provide connection density of 1 million devices per square kilometer. In case of URLLC, the main requirements include 0.5 ms end-to-end latency and reliability above 99.999% [16–18]. By using NOMA scheme, for mMTC and URLLC applications, the number of user connections can be increased by 5 and 9 times, respectively [18]. Also, according to [19], NOMA has been shown to be more spectral-efficient by 30% for downlink and 100% for uplink in eMBB when compared to OMA. Therefore, NOMA has been recognized as a strong candidate among all MA techniques since it has essential features to overcome challenges in counterpart OMA and achieve the requirements of next mobile communication systems [20–22]. The superiority of NOMA over OMA can be remarked as follows:

(i) Spectral efficiency and throughput: in OMA, such as in OFDMA, a specific frequency resource is assigned to each user even it experiences a good or bad channel condition; thus the overall system suffers from low spectral efficiency and throughput. In the contrary, in NOMA the same frequency resource is assigned to multiple mobile users, with good and bad channel conditions, at the same time. Hence, the resource assigned for the weak user is also used by the strong user, and the interference can be mitigated through SIC processes at users' receivers. Therefore, the probability of having improved spectral efficiency and a high throughput will be considerably increased as depicted in Figure 1.

(ii) User fairness, low latency, and massive connectivity: in OMA, for example in OFDMA with scheduling, the user with a good channel condition has a higher priority to be served while the user with a bad channel condition has to wait for access, which leads to a fairness problem and high latency. This approach can not support massive connectivity. However, NOMA can serve multiple users with different channel conditions simultaneously; therefore, it can provide improved user fairness, lower latency, and higher massive connectivity [20].

(iii) Compatibility: NOMA is also compatible with the current and future communication systems since it does not require significant modifications on the existing architecture. For example, NOMA has been included in third generation partnership project long-term evolution advanced (3GPP LTE Release 13) [23–29]. More detailed, in the standards, a downlink version of NOMA, multiuser superposition transmission (MUST), has been used [23]. MUST utilizes the superposition coding concept for a multiuser transmission in LTE-A systems. In 3GPP radio access network (RAN), while using MUST, the deployment scenarios, evaluation methodologies, and candidate NOMA scheme have been investigated in [24–26], respectively. Then, system level performance and link level performance of NOMA have been evaluated in [27, 28], respectively. Next, 3GPP LTE Release 14 has been proposed [29], in which intracell interference is eliminated and hence LTE can support downlink intracell multiuser superposition transmission. Also, NOMA, known as layered division multiplexing (LDM), is used in the future digital TV standard, ATSC 3.0 [30]. Moreover, the standardization study of NOMA schemes for 5G New Radio (NR) continues within 3GPP LTE Release 15 [31]. Agreed objectives in Release 15 can be summarized as follows: (1) transmitter side signal processing schemes for NOMA, such as modulation and symbol level processing, coded bit level processing, and symbol to resource element mapping; (2) receivers for NOMA, such as minimum mean-square error (MMSE) receiver, SIC and/or parallel interference cancellation (PIC) receiver, joint detection type receivers, and complexity of the receivers; (3) NOMA procedures, such as uplink transmission detection, link adaptation MA, synchronous and asynchronous operation, and adaptation between OMA and NOMA; (4) link and system level performance evaluation or analysis for NOMA, such as traffic model and deployment scenarios of eMBB, mMTC and URLLC, coverage, latency, and signaling overhead.

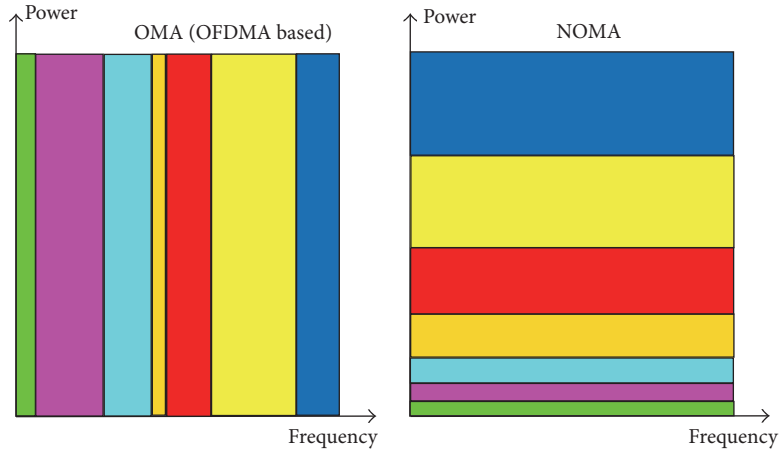

Figure 1: A pictorial comparison of OMA and NOMA.

In other words, the insufficient performance of OMA makes it inapplicable and unsuitable to provide the features needed to be met by the future generations of wireless communication systems. Consequently, researchers suggest NOMA as a strong candidate as an MA technique for next generations [32]. Although NOMA has many features that may support next generations, it has some limitations that should be addressed in order to exploit its full advantage set. Those limitations can be pointed out as follows. In NOMA, since each user requires to decode the signals of some users before decoding its own signal, the receiver computational complexity will be increased when compared to OMA, leading to a longer delay. Moreover, information of channel gains of all users should be fed back to the base station (BS), but this results in a significant channel state information (CSI) feedback overhead. Furthermore, if any errors occur during SIC processes at any user, then the error probability of successive decoding will be increased. As a result, the number of users should be reduced to avoid such error propagation. Another reason for restricting the number of users is that considerable channel gain differences among users with different channel conditions are needed to have a better network performance.

This paper, written in a tutorial name, focuses on NOMA technique, along with its usage in MIMO and cooperative scenarios. Practice implementation aspects are also detailed. Besides, an overview about the standardizations of NOMA in 3GPP LTE and application in the 5G scenarios is provided. In addition, unlike previous studies, this paper includes performance analyses of MIMO-NOMA and cooperative NOMA scenarios to make the NOMA concept more understandable by researchers. The remainder of this paper is organized as follows. Basic concepts of NOMA, in both downlink and uplink networks, are given in Section 2. In Sections 3 and 4, MIMO-NOMA and cooperative NOMA are described, respectively. Practical implementation challenges of NOMA are detailed in Section 5. The paper is concluded in Section 6.

## 2. Basic Concepts of NOMA

In this section, an overview of NOMA in downlink and uplink networks is introduced through signal-to-interference-and-noise ratio (SINR) and sum rate analyses. Then, high signal-to-noise ratio (SNR) analysis has been conducted in order to compare the performances of OMA and NOMA techniques.

*2.1. Downlink NOMA Network.* At the transmitter side of downlink NOMA network, as shown in Figure 2, the BS transmits the combined signal, which is a superposition of the desired signals of multiple users with different allocated power coefficients, to all mobile users. At the receiver of each user, SIC process is assumed to be performed successively until user's signal is recovered. Power coefficients of users are allocated according to their channel conditions, in an inversely proportional manner. The user with a bad channel condition is allocated higher transmission power than the one which has a good channel condition. Thus, since the user with the highest transmission power considers the signals of other users as noise, it recovers its signal immediately without performing any SIC process. However, other users need to perform SIC processes. In SIC, each user's receiver first detects the signals that are stronger than its own desired signal. Next, those signals are subtracted from the received signal and this process continues until the related user's own signal is determined. Finally, each user decodes its own signal by treating other users with lower power coefficients as noise. The transmitted signal at the BS can be written as follows:

$$s = \sum_{i=1}^{L} \sqrt{a_i P_s} x_i, \quad (1)$$

where $x_i$ is the information of user $i$ ($U_i$) with unit energy. $P_s$ is the transmission power at the BS and $a_i$ is the power coefficient allocated for user $i$ subjected to $\sum_{i=1}^{L} a_i = 1$ and $a_1 \geq a_2 \geq \cdots \geq a_L$ since without loss of generality the channel gains are assumed to be ordered as $|h_1|^2 \leq |h_2|^2 \leq \cdots \leq |h_L|^2$, where $h_l$ is the channel coefficient of $l$th user, based on NOMA concept. The received signal at $l$th user can be expressed as follows:

$$y_l = h_l s + n_l = h_l \sum_{i=1}^{L} \sqrt{a_i P_s} x_i + n_l, \quad (2)$$

where $n_l$ is zero mean complex additive Gaussian noise with a variance of $\sigma^2$; that is, $n_l \sim CN(0, \sigma^2)$.

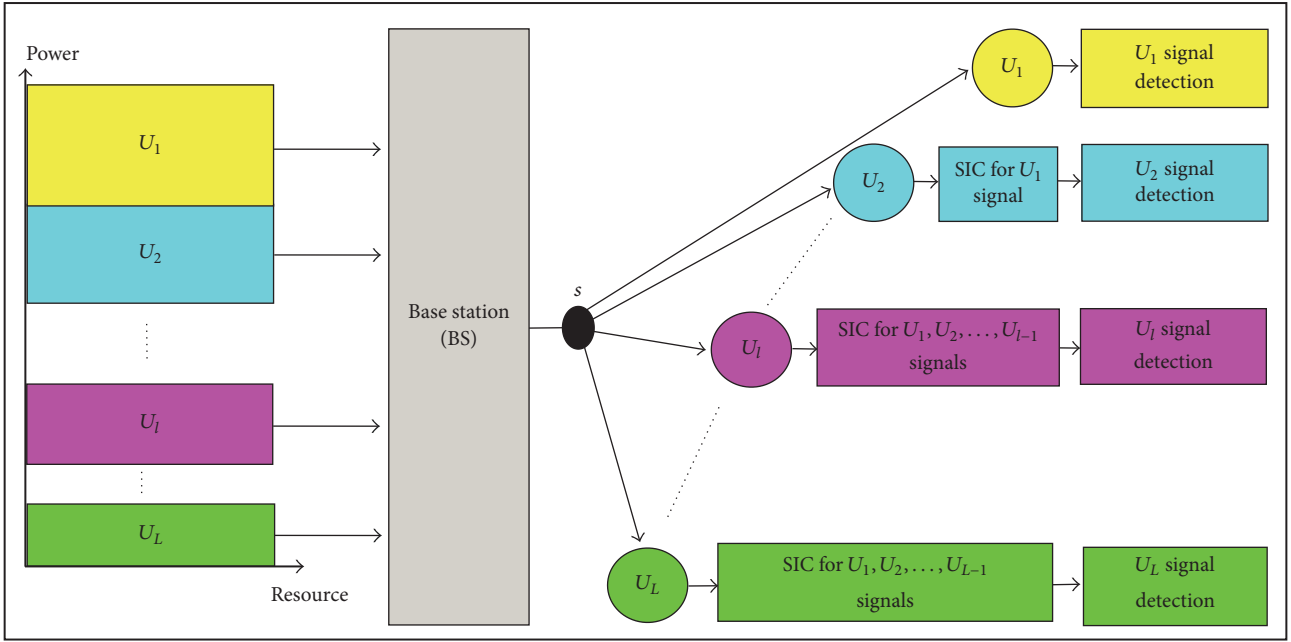

Figure 2: Downlink NOMA network.

*2.1.1. SINR Analysis.* By using (2), the instantaneous SINR of the $l$th user to detect the $j$th user, $j \leq l$, with $j \neq L$ can be written as follows:

$$\text{SINR}_{j \to l} = \frac{a_j \gamma |h_l|^2}{\gamma |h_l|^2 \sum_{i=j+1}^{L} a_i + 1}, \quad (3)$$

where $\gamma = P_s/\sigma^2$ denotes the SNR. In order to find the desired information of the $l$th user, SIC processes will be implemented for the signal of user $j \leq l$. Thus, the SINR of $l$th user can be given by

$$\text{SINR}_l = \frac{a_l \gamma |h_l|^2}{\gamma |h_l|^2 \sum_{i=l+1}^{L} a_i + 1}. \quad (4)$$

Then, the SINR of the $L$th user is expressed as

$$\text{SINR}_L = a_L \gamma |h_L|^2. \quad (5)$$

*2.1.2. Sum Rate Analysis.* After finding the SINR expressions of downlink NOMA, the sum rate analysis can easily be done. The downlink NOMA achievable data rate of $l$th user can be expressed as

$$R_l^{\text{NOMA-d}} = \log_2 (1 + \text{SINR}_l)$$
$$= \log_2 \left( 1 + \frac{a_l \gamma |h_l|^2}{\gamma |h_l|^2 \sum_{i=l+1}^{L} a_i + 1} \right). \quad (6)$$

Therefore, the sum rate of downlink NOMA can be written as

$$R_{\text{sum}}^{\text{NOMA-d}} = \sum_{l=1}^{L} \log_2 (1 + \text{SINR}_l)$$
$$= \sum_{l=1}^{L-1} \log_2 \left( 1 + \frac{a_l \gamma |h_l|^2}{\gamma |h_l|^2 \sum_{i=l+1}^{L} a_i + 1} \right)$$
$$+ \log_2 \left( 1 + a_L \gamma |h_L|^2 \right)$$

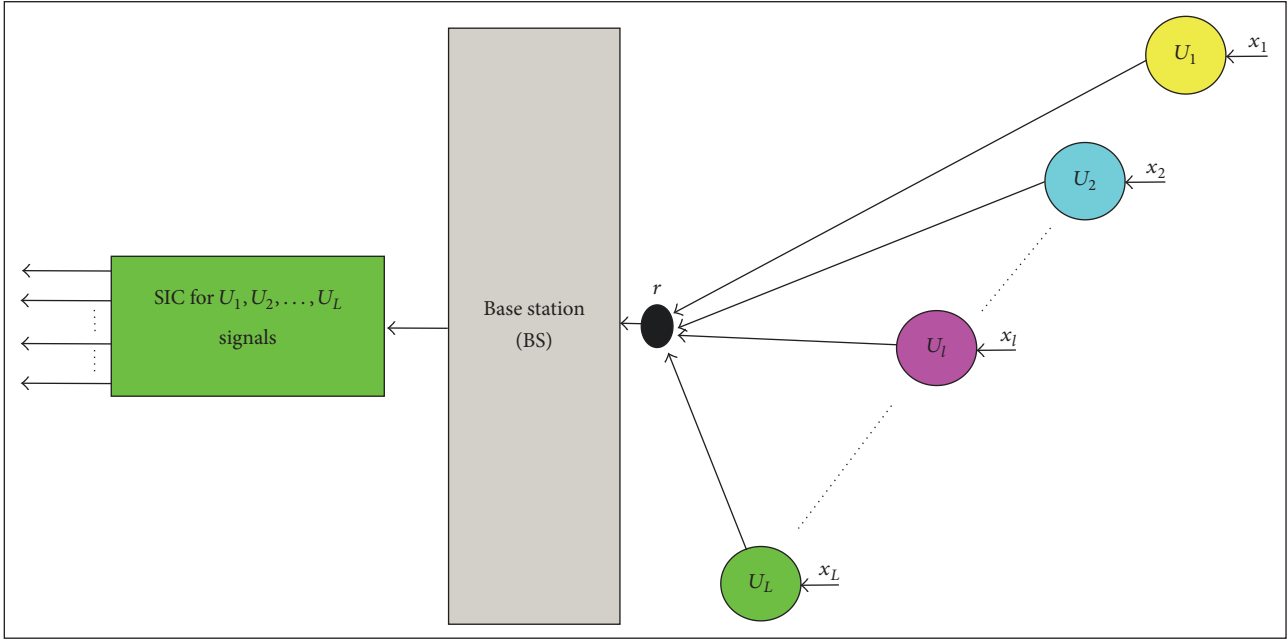

Figure 3: Uplink NOMA network.

$$= \sum_{l=1}^{L-1} \log_2 \left( 1 + \frac{a_l}{\sum_{i=l+1}^{L} a_i + 1/\gamma |h_l|^2} \right) + \log_2 \left( 1 + a_L \gamma |h_L|^2 \right). \tag{7}$$

In order to figure out whether NOMA techniques outperform OMA techniques, we conduct a high SNR analysis. Thus, at high SNR, that is, $\gamma \to \infty$, the sum rate of downlink NOMA becomes

$$R_{\text{sum}}^{\text{NOMA-d}} \approx \sum_{l=1}^{L-1} \log_2 \left( 1 + \frac{a_l}{\sum_{i=l+1}^{L} a_i} \right) + \log_2 \left( \gamma |h_L|^2 \right)$$
$$\approx \log_2 \left( \gamma |h_L|^2 \right). \tag{8}$$

### 2.2. Uplink NOMA Network.

In uplink NOMA network, as depicted in Figure 3, each mobile user transmits its signal to the BS. At the BS, SIC iterations are carried out in order to detect the signals of mobile users. By assuming that downlink and uplink channels are reciprocal and the BS transmits power allocation coefficients to mobile users, the received signal at the BS for synchronous uplink NOMA can be expressed as

$$r = \sum_{i=1}^{L} h_i \sqrt{a_i P} x_i + n, \tag{9}$$

where $h_i$ is the channel coefficient of the $i$th user, $P$ is the maximum transmission power assumed to be common for all users, and $n$ is zero mean complex additive Gaussian noise with a variance of $\sigma^2$; that is, $n \sim \text{CN}(0, \sigma^2)$.

#### 2.2.1. SINR Analysis.
The BS decodes the signals of users orderly according to power coefficients of users, and then the SINR for $l$th user $l \neq 1$ can be given by [33]

$$\text{SINR}_l = \frac{a_l \gamma |h_l|^2}{\gamma \sum_{i=1}^{l-1} a_i |h_i|^2 + 1}, \tag{10}$$

where $\gamma = P/\sigma^2$. Next, the SINR for the first user is expressed as

$$\text{SINR}_1 = a_1 \gamma |h_1|^2. \tag{11}$$

#### 2.2.2. Sum Rate Analysis.
The sum rate of uplink NOMA can be written as

$$\begin{aligned} R_{\text{sum}}^{\text{NOMA-u}} &= \sum_{l=1}^{L} \log_2 \left( 1 + \text{SINR}_l \right) \\ &= \log_2 \left( 1 + a_1 \gamma |h_1|^2 \right) \\ &\quad + \sum_{l=2}^{L} \log_2 \left( 1 + \frac{a_l \gamma |h_l|^2}{\gamma \sum_{i=1}^{l-1} a_i |h_i|^2 + 1} \right) \\ &= \log_2 \left( 1 + \gamma \sum_{l=1}^{L} a_l |h_l|^2 \right). \end{aligned} \tag{12}$$

When $\gamma \to \infty$, the sum rate of uplink NOMA becomes

$$R_{\text{sum}}^{\text{NOMA-u}} \approx \log_2 \left( \gamma \sum_{l=1}^{L} |h_l|^2 \right). \tag{13}$$

### 2.3. Comparing NOMA and OMA.

The achievable data rate of the $l$th user of OMA for both uplink and downlink can be expressed as [33]

$$R_l^{\text{OMA}} = \alpha_l \log_2\left(1 + \frac{\beta_l \gamma |h_l|^2}{\alpha_l}\right), \tag{14}$$

where $\beta_l$ and $\alpha_l$ are the power coefficient and the parameter related to the specific resource of $U_l$, respectively. And then, the sum rate of OMA is written as

$$R_{\text{sum}}^{\text{OMA}} = \sum_{l=1}^{L} \alpha_l \log_2\left(1 + \frac{\beta_l \gamma |h_l|^2}{\alpha_l}\right). \tag{15}$$

For OMA, for example, FDMA, total bandwidth resource and power are shared among the users equally; then using $\alpha_l = \beta_l = 1/L$ the sum rate can be written as

$$R_{\text{sum}}^{\text{OMA}} = \sum_{l=1}^{L} \frac{1}{L} \log_2\left(1 + \gamma |h_l|^2\right). \tag{16}$$

When $\gamma \to \infty$, the sum rate of OMA becomes

$$R_{\text{sum}}^{\text{OMA}} \approx \sum_{l=1}^{L} \frac{1}{L} \log_2\left(\gamma |h_l|^2\right). \tag{17}$$

Using $|h_1|^2 \leq |h_2|^2 \leq \cdots \leq |h_L|^2$,

$$R_{\text{sum}}^{\text{OMA}} \approx \sum_{l=1}^{L} \frac{1}{L} \log_2\left(\gamma |h_l|^2\right) \leq \sum_{l=1}^{L} \frac{1}{L} \log_2\left(\gamma |h_L|^2\right) \\ = \log_2\left(\gamma |h_L|^2\right) \approx R_{\text{sum}}^{\text{NOMA-d}}. \tag{18}$$

Hence, we conclude $R_{\text{sum}}^{\text{OMA}} \leq R_{\text{sum}}^{\text{NOMA-d}}$.

For the sake of simplicity, sum rates of uplink NOMA and OMA can be compared for two users. Then, using (13) and (17) the sum rate of uplink NOMA and OMA at high SNR can be expressed, respectively, as

$$R_{\text{sum}}^{\text{NOMA-u}} \approx \log_2\left(\gamma |h_1|^2 + \gamma |h_2|^2\right), \tag{19}$$

$$R_{\text{sum}}^{\text{OMA}} \approx \frac{1}{2}\log_2\left(\gamma |h_1|^2\right) + \frac{1}{2}\log_2\left(\gamma |h_2|^2\right) \\ \leq \log_2\left(\gamma |h_2|^2\right). \tag{20}$$

From (19) and (20), we notice $R_{\text{sum}}^{\text{OMA}} \leq R_{\text{sum}}^{\text{NOMA-u}}$.

Figure 4 shows that NOMA outperforms OMA in terms of sum rate in both downlink and uplink of two user networks using (7), (12), and (16).

## 3. MIMO-NOMA

MIMO technologies have a significant capability of increasing capacity as well as improving error probability of wireless communication systems [34]. To take advantage of MIMO schemes, researchers have investigated the performance of NOMA over MIMO networks [35]. Many works have been studying the superiority of MIMO-NOMA over MIMO-OMA in terms of sum rate and ergodic sum rate under different conditions and several constrictions [36–39]. Specifically, in [36], the maximization problem of ergodic sum rate for two-user MIMO-NOMA system over Rayleigh fading channels is discussed. With the need of partial CSI at the BS and under some limitations on both total transmission power and the minimum rate for the user with bad channel condition, the optimal power allocation algorithm with a lower complexity to maximize the ergodic capacity is proposed. However, in order to achieve a balance between the maximum number of mobile users and the optimal achievable sum rate in MIMO-NOMA systems, sum rate has been represented through two ways. The first approach targets the optimization of power partition among the user clusters [37]. Another approach is to group the users in different clusters such that each cluster can be allocated with orthogonal spectrum resources according to the selected user grouping algorithm [38]. Furthermore, in [37] performances of two users per cluster schemes have been studied for both MIMO-NOMA and MIMO-OMA over Rayleigh fading channels. In addition, in accordance with specified power split, the dominance of NOMA over OMA has been shown in terms of sum channel and ergodic capacities.

On the other side, the authors in [38] have examined the performance of MIMO-NOMA system, in which multiple users are arranged into a cluster. An analytical comparison has been provided between MIMO-NOMA and MIMO-OMA, and then it is shown that NOMA outperforms OMA in terms of sum channel and ergodic capacities in case of multiple antennas. Moreover, since the number of users per cluster is inversely proportional to the achievable sum rate and the trade-off between the number of admitted users and achieved sum rate has to be taken into account (which restricts the system performance), a user admission scheme, which maximizes the number of users per cluster based on their SINR thresholds, is proposed. Although the optimum performance is achieved in terms of the number of admitted users and the sum rate when the SINR thresholds of all users are equal, even when they are different good results are obtained. In addition, a low complexity of the proposed scheme is linearly proportional to the number of users per cluster. In [39], the performance of downlink MIMO-NOMA network for a simple case of two users, that is, one cluster, is introduced. In this case, MIMO-NOMA provides a better performance than MIMO-OMA in terms of both the sum rate and ergodic sum rate. Also, it is shown that for a more practical case of multiple users, with two users allocated into a cluster and sharing the same transmit beamforming vector, where ZF precoding and signal alignment are employed at the BS and the users of the same cluster, respectively, the same result still holds.

Antenna selection techniques have also been recognized as a powerful solution that can be applied to MIMO systems in order to avoid the adverse effects of using multiple antennas simultaneously. These effects include hardware complexity, redundant power consumption, and high cost. Meanwhile diversity advantages that can be achieved from MIMO systems are still maintained [40]. Several works apply antenna selection techniques in MIMO-NOMA as they have already been developed for MIMO-OMA systems. But the gains can not be easily replicated since there is a heavy interuser interference in MIMO-NOMA networks, dissimilar from those in MIMO-OMA networks, in which information is transmitted in an interference-free manner. Consequently, there are a few works that challenged the antenna selection problem [41–43]. In [41], the sum rate performance for downlink multiple input single output- (MISO-) NOMA system is investigated with the help of transmit antenna selection (TAS) at the BS, where the transmitter of the BS and the receiver of each mobile user are equipped with multiantenna and single antenna, respectively. Basically, in TAS-OMA scheme, the best antenna at the BS offering the highest SINR is selected. However in the proposed TAS-NOMA scheme in [41], the best antenna at the BS providing the maximum sum rate is chosen. In addition to using an efficient TAS scheme, user scheduling algorithm is applied in two user massive MIMO-NOMA system in order to maximize the achievable sum rate in [42] for two scenarios, namely, the single-band two users and the multiband multiuser. In the first scenario, an efficient search algorithm is suggested. This algorithm aims to choose the antennas providing the highest channel gains in such a way that the desired antennas are only searched from specified finite candidate set, which are useful to the concerned users. On the other hand, in the second scenario, a joint user and antenna contribution algorithm is proposed. In particular, this algorithm manipulates the ratio of channel gain specified by a certain antenna-user pair to the total channel gain, and hence antenna-user pair offering the highest contribution to the total channel gain is selected. Moreover, an efficient search algorithm provides a better trade-off between system performance and complexity, rather than a joint antenna and user contribution algorithm. Unfortunately, neither the authors of [41] nor the authors of [42] have studied the system performance analytically. In [43], the maximization of the average sum rate of two-user NOMA system, in which the BS and mobile users are equipped with multiantenna, is discussed through two computationally effective joint antenna selection algorithms; the max-min-max and the max-max-max algorithms. However, the instantaneous channel gain of the user with a bad channel condition is improved in max-min-max antenna selection scheme while max-max-max algorithm is the solution for the user with a good channel condition. Furthermore, asymptotic closed-form expressions of the average sum rates are evaluated for both proposed algorithms. Moreover, it is verified that better user fairness can be achieved by the max-min-max algorithm while larger sum rate can be obtained by the max-max-max algorithm.

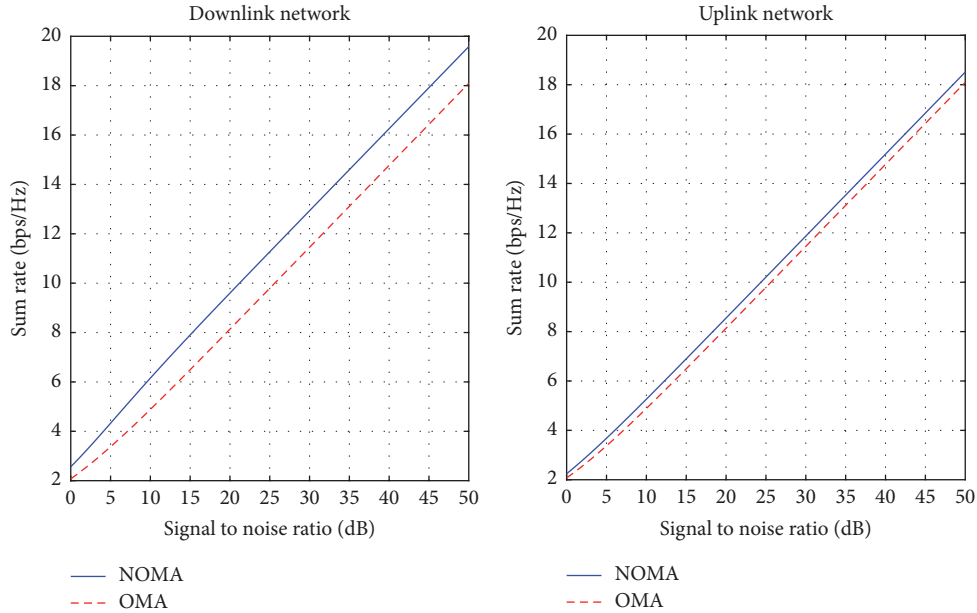

FIGURE 4: Sum rate of NOMA and OMA in both downlink and uplink networks with $a_1 = 0.6$, $a_2 = 0.4$, $|h_1|^2 = 0$ dB, and $|h_2|^2 = 20$ dB.

Multicast beamforming can also be introduced as a technique that can be employed in MIMO schemes since it offers a better sum capacity performance even for multiple users. However, it can be applied in different ways. One approach is based on a single beam that can be used by all users; hence all users receive this common signal [44]. Another approach is to use multiple beams that can be utilized by many groups of users; that is, each group receives a different signal [45]. The following works have studied beamforming in MIMO-NOMA systems. In [46], multiuser beamforming in downlink MIMO-NOMA system is proposed. Particularly, a pair of users can share the same beam. Since the proposed beam can be only shared by two users with different channel qualities, it is probable to easily apply clustering and power allocation algorithms to maximize the sum capacity and to decrease the intercluster and interuser interferences. In [47], performance of multicast beamforming, when the beam is used to serve many users per cluster by sharing a common signal, is investigated with superposition coding for a downlink MISO-NOMA network in a simple scenario of two users.

Principally, the transmitter of the BS has multiantenna and its information stream is based on multiresolution broadcast concept, in which only low priority signal is sent to the user that is far away from the BS, that is, user with a bad channel quality. Both signals of high priority and low priority are transmitted to the user near to BS, that is, user with good channel quality. Furthermore, with superposition coding a minimum power beamforming problem has been developed in order to find the beamforming vectors and the powers for both users. Moreover, under the considered optimization condition and the given normalized beamforming vectors (which are founded by an iterative algorithm), the closed-form expression for optimal power allocation is easily obtained. In [48], random beamforming is carried out at the BS of a downlink MIMO-NOMA network. In the system model, each beam is assumed to be used by all the users in one cluster and all beams have similar transmission power allocations. Moreover, a spatial filter is suggested to be used in order to diminish the intercluster and interbeam interferences. Fractional frequency reuse concept, in which users with different channel conditions can accommodate many reuse factors, is proposed in order to improve the power allocation among multiple beams. In [49], interference minimization and capacity maximization for downlink multiuser MIMO-NOMA system are introduced, in which the number of receive antennas of mobile user is larger than the number of transmit antennas of the BS. Zero-forcing beamforming technique is suggested to reduce the intercluster interference, especially when distinctive channel quality users is assumed. In addition, dynamic power allocation and user-cluster algorithms have been proposed not only to achieve maximum throughput, but also to minimize the interference.

There are many research works investigating resource allocation problem in terms of maximization of the sum rate in case of perfect CSI [50–52]. Specifically, in [50] sum rate optimization problem of two-user MIMO-NOMA network, that is, two users in one cluster in which different precoders are implemented, has been introduced under the constraint of transmission power at the BS and the minimum transmission rate limitation of the user with bad channel condition. In [51], the sum rate maximization problem for downlink MISO-NOMA system is investigated. However, the transmitted signal for each mobile user is weighted with a complex vector. Moreover, for the sake of avoiding the high computational complexity related to nonconvex optimization problem, minorization-maximization method is suggested as an approximation. The key idea of minorization-maximization algorithm is to design the complex weighting vectors in such a way that the total throughput of the system is maximized, for a given order of users; that is, perfect CSI is assumed. In [52], a downlink MIMO-NOMA system, where perfect CSI available at all nodes is assumed and with different beams, BS broadcasts precoded signals to all mobile users; that is, each beam serves several users. However, there are three proposed algorithms combined in order to maximize the sum rate. The first one is where weighted sum rate maximization proposes to design a special beamforming matrix of each beam benefiting from all CSI at the BS. The second algorithm is where user scheduling aims to have super SIC at the receiver of each mobile user. Thus, to take full benefits of SIC, differences in channel gains per cluster should be significant and the channel correlation between mobile users has to be large. The final one is where fixed power allocation targets optimization, offering not only a higher sum rate, but also convenient performance for the user with bad channel quality. In [53], the optimal power allocation method, in order to maximize the sum rate of two-user MIMO-NOMA with a layered transmission scheme under a maximum transmission power constraint for each mobile user, is investigated. Basically, by using the layered transmission, each mobile user performs sequence by sequence decoding signals throughout SIC, yielding much lower decoding complexity when compared to the case with nonlayered transmission. Moreover, the closed-form expression for the average sum rate and its bounds in both cases of perfect CSI and partial CSI are obtained. Also, it is shown that the average sum rate is linearly proportional to the number of antennas. In [54], a comprehensive resource allocation method for multiuser downlink MIMO-NOMA system including beamforming and user selection is proposed, yielding low computational complexity and high performance in cases of full and partial CSI. However, resource allocation has been expressed in terms of the maximum sum rate and the minimum of maximum outage probability (OP) for full CSI and partial CSI, respectively. Outage behavior for both downlink and uplink networks in MIMO-NOMA framework with integrated alignment principles is investigated in a single cell [55] and multicell [56, 57], respectively. Furthermore, an appropriate trade-off between fairness and throughput has been achieved by applying two strategies of power allocation methods. The fixed power allocation strategy realizes different QoS requirements. On the other hand cognitive radio inspired power allocation strategy verifies that QoS requirements of the user are achieved immediately. In addition, exact and asymptotic expressions of the system OP have been derived. In [58], the power minimization problem for downlink MIMO-NOMA networks under full CSI and channel distribution information scenarios are studied. In [59], linear beamformers, that is, precoders that provide a larger total sum throughput also improving throughput of the user with bad quality channel, are designed; meanwhile QoS specification requirements are satisfied. Also, it is shown that the maximum number of users per cluster that realizes a higher NOMA performance is achieved at larger distinctive channel gains.

Moreover, since massive MIMO technologies can ensure bountiful antenna diversity at a lower cost [4], many works have discussed performance of NOMA over massive MIMO. For instance, in [60], massive MIMO-NOMA system, where the number of the transmit antennas at the BS is significantly larger than the number of users, is studied with limited feedback. Also, the exact expressions of the OP and the diversity order are obtained for the scenarios of perfect order of users and one bit feedback, respectively. In [61], the scheme based on interleave division multiple access and iterative data-aided channel estimation is presented in order to solve the reliability problem of multiuser massive MIMO-NOMA system with imperfect CSI available at the BS. In [62], the achievable rate in massive MIMO-NOMA systems and iterative data-aided channel estimation receiver, in which partially decoded information is required to get a better channel estimation, are investigated through applying two pilot schemes: orthogonal pilot and superimposed pilot. However, pilots in the orthogonal pilot scheme occupy time/frequency slots while they are superimposed with information in superimposed pilot one. Moreover, it is shown that the greatest part of pilot power in superimposed pilot scheme seems to be zero in the case when Gaussian signal prohibits overhead power and rate loss that may be resulted through using pilot. Consequently, with code maximization superimposed scheme has a superior performance over orthogonal one under higher mobility and larger number of mobile users. Different from massive MIMO, in [63] performance of massive access MIMO systems, in which number of users is larger than the number of antennas employed at the BS, is studied. Low-complexity Gaussian message specially passing iterative detection algorithm is used and both its mean and variance precisely converge with high speed to those concerned with the minimum mean square error multiuser detection in [64].

In addition, NOMA has been proposed as a candidate MA scheme integrated with beamspace MIMO in mmWave communication systems, satisfying massive connectivity, where the number of mobile users is much greater than the number of radio frequency chains, and obtaining a better performance in terms of spectrum and energy efficiency [65]. Furthermore, a precoding scheme designed on zero-forcing (ZF) concept has been suggested in order to reduce the interbeam interference. Moreover, iterative optimization algorithm with dynamic power allocation scheme is proposed to obtain a higher sum rate and lower complexity. In [66], the optimization problem of energy efficiency for MIMO-NOMA systems with imperfect CSI at the BS over Rayleigh fading channels is studied under specified limitations on total transmission power and minimum sum rate of the user of bad channel condition. However, two-user scheduling schemes and power allocation scheme are presented in [67] in order to maximize the energy efficiency. The user scheduling schemes depend on the signal space alignment; while one of them effectively deals with the multiple interference, the other one maximizes the multicollinearity among users. On the other hand, power allocation scheme uses a sequential convex approximation that roughly equalizes the nonconvex problem by a set of convex problems iteratively, that is, in each iteration nonconvex constraints are modified into their approximations in inner convex. Also, it is shown that higher energy efficiency is obtained when lower power is transmitted and a higher sum rate of center users is obtained when maximum multicollinearity scheme is employed.

Many other problems have been investigated in MIMO-NOMA systems. For example, in [68, 69], QoS optimization problem is proposed for two-user MISO-NOMA system. In particular, closed-form expressions of optimal precoding vectors over flat fading channels, are achieved by applying the Lagrange duality and an iterative method in [68] and [69], respectively.

As mentioned before, NOMA promises to satisfy the need of IoT, in which many users require to be served rapidly for small packet transmissions. Consequently, the literature tends to study performance of MIMO-NOMA for IoT. For instance, in [70] a MIMO-NOMA downlink network where one transmitter sending information to two users is considered. However, one user has a low data rate, that is, small packet transmission, while the second user has a higher rate. Particularly, outage performance in case of using precoding and power allocation method is investigated. Also, it is shown that the potential of NOMA is apparent even when channel qualities of users are similar.

Most current works of MIMO-NOMA focus on sum rate and capacity optimization problems. However, performance of symbol error rate (SER) for wireless communication systems is also very substantial. In [71], SER performance using the minimum Euclidean distance precoding scheme in MIMO-NOMA networks is studied. For simple transmission case, two-user $2 \times 2$ MIMO-NOMA is investigated. However, to facilitate realization of practical case of multiuser MIMO-NOMA network, two-user pairing algorithms are applied.

In order to demonstrate the significant performance of MIMO-NOMA systems in terms of both OP and sum rate, as well as its superiority over MIMO-OMA, a special case, performance of single input multiple output- (SIMO-) NOMA network based on maximal ratio combining (MRC) diversity technique in terms of both OP and ergodic sum rate is investigated in the following section. Moreover, closed-form expression of OP and bounds of ergodic sum rate are derived.

### 3.1. Performance Analysis of SIMO-NOMA.

This network includes a BS and $L$ mobile users as shown in Figure 5. The transmitter of BS is equipped with a single antenna and the receiver of each mobile user is equipped with $N_r$ antennas. The received signal at the $l$th user after applying MRC can be written as follows:

$$r_l = \|\mathbf{h_l}\| \sum_{i=1}^{L} \sqrt{a_i P_s} x_i + \frac{\mathbf{h_l^H}}{\|\mathbf{h_l}\|} \mathbf{n_l}, \qquad (21)$$

where $\mathbf{h_l}$ is $N_r \times 1$ fading channel coefficient vector between the BS and $l$th user and without loss of generality and due to NOMA concept they are sorted in ascending way; that is, $\|\mathbf{h_1}\|^2 \leq \|\mathbf{h_2}\|^2 \leq \cdots \leq \|\mathbf{h_L}\|^2$, and $\mathbf{n_l}$ is $N_r \times 1$ zero mean complex additive Gaussian noise with $E[\mathbf{n_l}\mathbf{n_l^H}] = \mathbf{I}_{N_r}\sigma_l^2$ at the $l$th user, where $E[\cdot]$, $(\cdot)^H$, and $\mathbf{I}_r$ denote the expectation operator, Hermitian transpose, and identity matrix of order $r$, respectively, and $\sigma_l^2 = \sigma^2$ is the variance of $\mathbf{n_l}$ per dimension. From (21), instantaneous SINR for $l$th user to detect $j$th user, $j \leq l$, with $j \neq L$ can be expressed as follows:

$$\text{SINR}_{j \to l} = \frac{a_j \gamma \|\mathbf{h_l}\|^2}{\gamma \|\mathbf{h_l}\|^2 \sum_{i=j+1}^{L} a_i + 1}. \qquad (22)$$

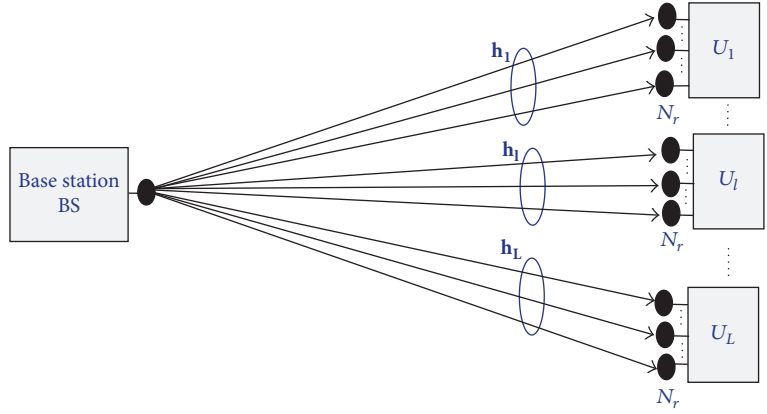

Figure 5: System model of the downlink SIMO-NOMA.

Now, nonordered channel gains for MRC can be given as follows:

$$\|\tilde{\mathbf{h}}_\mathbf{l}\|^2 = \sum_{i=1}^{N_r} |h_{l,i}|^2, \quad l = 1, 2, \ldots, L, \tag{23}$$

where $h_{l,i}$ denotes the channel coefficient between the BS and $i$th antenna of the $l$th user and are independent and identically distributed (i.i.d.) Nakagami-$m$ random variables. By the help of the series expansion of incomplete Gamma function [72, eq. (8.352.6)], the cumulative distribution function (CDF) and probability density function (PDF) of Gamma random variable $X$, square of Nakagami-$m$ random variable can be defined as follows:

$$F_X(x) = \frac{\gamma(m, mx/\Omega)}{\Gamma(m)} = 1 - e^{-mx/\Omega} \sum_{k=0}^{m-1} \left(\frac{mx}{\Omega}\right)^k \frac{1}{k!},$$

$$f_X(x) = \left(\frac{m}{\Omega}\right)^m \frac{x^{m-1}}{\Gamma(m)} e^{-mx/\Omega}, \tag{24}$$

where $\gamma(\cdot,\cdot)$ and $\Gamma(\cdot)$ are the lower incomplete Gamma function given by [72, eq. (8.350.1)] and the Gamma function given by [72, eq. (8.310.1)], respectively. $m$ is parameter of Nakagami-$m$ distribution, and $\Omega = E[|X|^2]$. With the help of the highest order statistics [73], we can write CDF of nonordered $\|\tilde{\mathbf{h}}_\mathbf{l}\|^2$ as follows:

$$F_{\|\tilde{\mathbf{h}}_\mathbf{l}\|^2}(x) = \frac{\gamma(mN_r, mx/\Omega)}{\Gamma(mN_r)}$$

$$= 1 - e^{-mx/\Omega} \sum_{s=0}^{mN_r-1} \left(\frac{mx}{\Omega}\right)^s \frac{1}{s!} \tag{25}$$

$$= \sum_{r=0}^{1} \sum_{s=0}^{r(mN_r-1)} (-1)^r \vartheta_s(r, mN_r) x^s e^{-rmx/\Omega},$$

where $\Omega = E[\|\tilde{\mathbf{h}}_\mathbf{l}\|^2]$ and $\vartheta_a(b, g_c)$ denotes multinomial coefficients which can be defined as [72, eq. (0.314)]

$$\vartheta_a(b, g_c) = \frac{1}{ad_0} \sum_{\rho=1}^{a} (\rho(b+1) - a) d_\rho \vartheta_{a-b}(b, g_c), \tag{26}$$

$$a \geq 1.$$

In (26), $d_\rho = (g_c/\Omega)^\rho/\rho!$, $\vartheta_0(b, g_c) = 1$, and $\vartheta_a(b, g_c) = 0$ if $\rho > g_c - 1$. Next, CDF of the ordered $\|\mathbf{h}_\mathbf{l}\|^2$ can be expressed as [74]

$$F_{\|\mathbf{h}_\mathbf{l}\|^2}(x) = \frac{L!}{(L-l)!(l-1)!} \sum_{t=0}^{L-l} \frac{(-1)^t}{l+t} \binom{L-l}{t}$$

$$\times \left[F_{\|\tilde{\mathbf{h}}_\mathbf{l}\|^2}(x)\right]^{l+t} = \frac{L!}{(L-l)!(l-1)!}$$

$$\cdot \sum_{t=0}^{L-l} \sum_{r=0}^{l+t} \sum_{s=0}^{r(mN_r-1)} \frac{(-1)^{t+r}}{l+t} \tag{27}$$

$$\cdot \binom{L-l}{t}\binom{l+t}{r} \vartheta_s(r, mN_r) x^s e^{-rmx/\Omega}.$$

*3.1.1. Outage Probability of SIMO-NOMA.* The OP of the $l$th user can be obtained as follows:

$$P_{\text{out},l} = \Pr\left(\text{SINR}_{j \to l} < \gamma_{\text{th}_j}\right)$$

$$= \Pr\left(\frac{a_j \gamma \|\mathbf{h}_\mathbf{l}\|^2}{\gamma \|\mathbf{h}_\mathbf{l}\|^2 \sum_{i=l+1}^{L} a_i + 1} < \gamma_{\text{th}_j}\right)$$

$$= \Pr\left(\|\mathbf{h}_\mathbf{l}\|^2 < \frac{\gamma_{\text{th}_j}}{\gamma\left(a_j - \gamma_{\text{th}_j} \sum_{i=l+1}^{L} a_i\right)}\right)$$

$$= \Pr\left(\|\mathbf{h_l}\|^2 < \eta_l^*\right) = F_{\|\mathbf{h_l}\|^2}\left(\eta_l^*\right) = \frac{L!}{(L-l)!(l-1)!}$$

$$\cdot \sum_{t=0}^{L-l} \sum_{r=0}^{l+t} \sum_{s=0}^{r(mN_r-1)} \frac{(-1)^{t+r}}{l+t}$$

$$\cdot \binom{L-l}{t} \binom{l+t}{r} \vartheta_s(r, mN_r) \eta_l^{*s} e^{-rm\eta_l^*/\Omega}, \tag{28}$$

where $\eta_l^* = \max[\eta_1, \eta_2, \ldots, \eta_l]$ with $\eta_j = \gamma_{th_j}/\gamma(a_j - \gamma_{th_j} \sum_{i=l+1}^{L} a_i)$. $\gamma_{th_j}$ denotes the threshold SINR of the $j$th user. Under the condition $a_j > \gamma_{th_j} \sum_{i=j+1}^{L} a_i$, the $l$th user can decode the $j$th user's signal successfully irrespective of the channel SNR.

*3.1.2. Ergodic Sum Rate Analysis of SIMO-NOMA.* Ergodic sum rate can be expressed as

$$R_{\text{sum}} = \sum_{l=1}^{L} E\left[\frac{1}{2}\log_2\left(1 + \text{SINR}_l\right)\right]$$

$$= \underbrace{\sum_{l=1}^{L-1} E\left[\frac{1}{2}\log_2\left(1 + \text{SINR}_l\right)\right]}_{R_{\overline{L}}} \tag{29}$$

$$+ \underbrace{E\left[\frac{1}{2}\log_2\left(1 + \text{SINR}_L\right)\right]}_{R_L}.$$

Then, $R_{\overline{L}}$ can be expressed as

$$R_{\overline{L}} = \sum_{l=1}^{L-1} E\left[\frac{1}{2}\log_2\left(1 + \frac{a_l \gamma \|\mathbf{h_l}\|^2}{\gamma \|\mathbf{h_l}\|^2 \sum_{i=l+1}^{L} a_i + 1}\right)\right]$$

$$= \sum_{l=1}^{L-1} E\left[\frac{1}{2}\log_2\left(1 + \frac{a_l}{\sum_{i=l+1}^{L} a_i + 1/\gamma \|\mathbf{h_l}\|^2}\right)\right]. \tag{30}$$

Due to computational difficulty of calculating the exact expression of the ergodic sum rate, and, for the sake of simplicity, we will apply high SNR analysis in order to find the upper and lower bounds related to ergodic sum rate. Thus, when $\gamma \to \infty$ in (30), then $R_{\overline{L}}^{\infty}$ can be given by

$$R_{\overline{L}}^{\infty} = \frac{1}{2} \sum_{l=1}^{L-1} \log_2\left(1 + \frac{a_l}{\sum_{i=l+1}^{L} a_i}\right). \tag{31}$$

Now, by using the identity $\int_0^{\infty} \ln(1+ay)f(y)dy = a\int_0^{\infty}((1-F(y))/(1+ay))dy$, $\log_b a = \ln a/\ln b$, $R_L$ can be written as

$$R_L = E\left[\frac{1}{2}\log_2\left(1 + a_L\gamma \|\mathbf{h_L}\|^2\right)\right]$$

$$= \frac{1}{2\ln 2} E\left[\ln\left(1 + a_L\gamma \|\mathbf{h_L}\|^2\right)\right]$$

$$= \frac{1}{2\ln 2} \int_0^{\infty} \ln\left(1 + \alpha_L\gamma x\right) f_{\|\mathbf{h_L}\|^2}(x) dx$$

$$= \frac{a_L\gamma}{2\ln 2} \int_0^{\infty} \frac{1 - F_{\|\mathbf{h_L}\|^2}(x)}{1 + a_L\gamma x} dx, \tag{32}$$

Simply, by using (27) $F_{\|\mathbf{h_L}\|^2}$ can be expressed as

$$F_{\|\mathbf{h_L}\|^2}(x)$$

$$= 1$$

$$+ \sum_{k=1}^{L} \sum_{n=0}^{k(mN_r-1)} \binom{L}{k} (-1)^k \vartheta_n(k, mN_r) x^n e^{-kmx/\Omega}. \tag{33}$$

By substituting (33) into (32),

$$R_L = \frac{a_L\gamma}{2\ln 2} \sum_{k=1}^{L} \sum_{n=0}^{k(mN_r-1)} \binom{L}{k} (-1)^{k+1} \vartheta_n(k, mN_r)$$

$$\cdot \underbrace{\int_0^{\infty} \frac{x^n e^{-kmx/\Omega}}{1 + a_L\gamma x} dx}_{I}. \tag{34}$$

By defining $u = a_L\gamma x$, $I$ can be written as follows:

$$I = \frac{1}{(a_L\gamma)^{n-1}} \int_0^{\infty} \frac{u^n e^{-kmu/a_L\gamma\Omega}}{1+u} du. \tag{35}$$

Using [74, (eq. 11)], as $\gamma \to \infty$, then $I$ can be approximated as

$$I \approx \xi = \begin{cases} \dfrac{\ln\left(a_L\gamma\Omega/mk\right)}{a_L\gamma}, & n = 0 \\ \dfrac{\Gamma(n)(\Omega/mk)^n}{a_L\gamma}, & n > 0. \end{cases} \tag{36}$$

By substituting (36) into (34), then $R_L^{\infty}$ can be given by

$$R_L^{\infty} = \frac{a_L\gamma}{2\ln 2} \sum_{k=1}^{L} \sum_{n=0}^{k(mN_r-1)} \binom{L}{k} (-1)^{k+1} \vartheta_n(k, mN_r) \xi. \tag{37}$$

Finally, by substituting (37) and (31) into (29), then asymptotic ergodic sum rate $R_{\text{sum}}^{\infty}$ can be expressed as

$$R_{\text{sum}}^{\infty}$$

$$= \frac{1}{2} \sum_{l=1}^{L-1} \log_2\left(1 + \frac{a_l}{\sum_{i=l+1}^{L} a_i}\right)$$

$$+ \frac{a_L\gamma}{2\ln 2} \sum_{k=1}^{L} \sum_{n=0}^{k(mN_r-1)} \binom{L}{k} (-1)^{k+1} \vartheta_n(k, mN_r) \xi. \tag{38}$$

*3.1.3. Numerical Results of SIMO-NOMA.* We consider two users and their average power factors that provide $\sum_{i=1}^{L} a_i = 1$ are selected as $a_1 = 0.6$ and $a_2 = 0.4$, respectively. Also, in order to make a comparison between the performances of conventional OMA and the proposed NOMA in terms of OP and ergodic sum rate over Nakagami-$m$ fading channels, SNR threshold value of conventional OMA $\gamma_{\text{th}}$, which verifies $(1/2) \sum_{i=1}^{L} \log_2(1 + \gamma_{\text{th}_i}) = (1/2)\log_2(1 + \gamma_{\text{th}})$, is used.

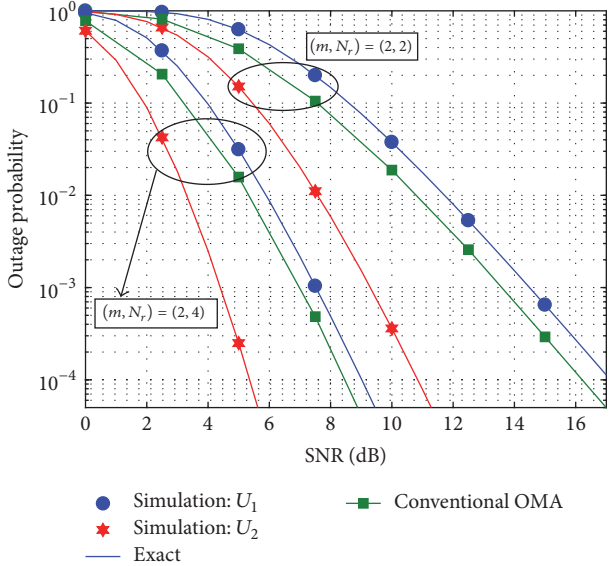

Figure 6: Outage probability of MIMO-NOMA system versus SNR for $L = 2$, $a_1 = 0.6$, $a_2 = 0.4$, $\gamma_{\text{th}_1} = 1$, $\gamma_{\text{th}_2} = 2$, and $\gamma_{\text{th}} = 5$.

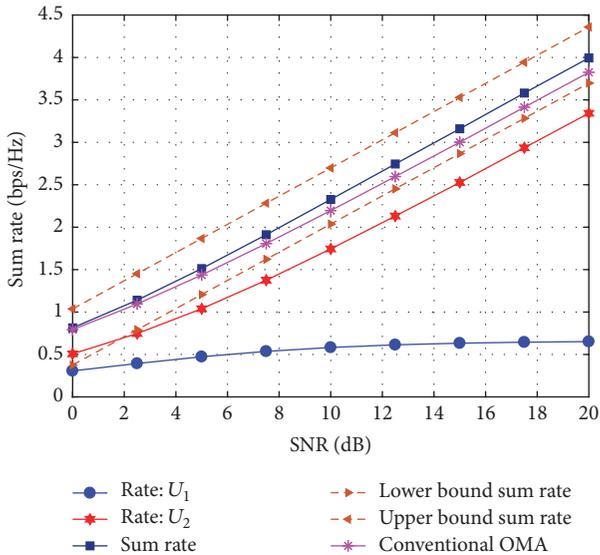

Figure 7: Ergodic sum rate of MIMO-NOMA system versus SNR for $L = 2$, $a_1 = 0.6$, $a_2 = 0.4$, $\gamma_{\text{th}_1} = 1$, $\gamma_{\text{th}_2} = 2$, $\gamma_{\text{th}} = 5$, and $(m, N_r) = (2, 2)$.

Figure 6 shows the outage probability versus the system SNR over different Nakagami $m$ parameters. In Figure 6, the simulations verify exact analytical results and a better outage performance at higher number of antennas is obtained.

Figure 7 depicts the ergodic sum rates of mobile users versus the system SNR. It is observed that ergodic rate for the first user is approximately constant over high SNR. This is due to high power allocation for the first user, such that it considers the signal of the second user as noise, while ergodic rate for the second user proportionally increases with SNR because of no interference with the first one. Figures 6 and 7 show that NOMA outperforms conventional OMA in terms of outage probability and ergodic sum rate, respectively.

## 4. Cooperative NOMA

Cooperative communication, where the transmission between the source and destination is maintained by the help of one or multiple relays, has received significant attention of researchers since it extends the coverage area and increases system capacity while reducing the performance deteriorating effects of multipath fading [75, 76]. In cooperative communication systems, relays transmit the received information signals to the related destinations by applying forwarding protocols, such as amplify-and-forward (AF) and decode-and-forward (DF). In addition, in the last decade, the relays can be fundamentally categorized as half-duplex (HD) and full-duplex (FD) according to relaying operation. Differing from HD, FD relay maintains the data reception and transmission process simultaneously in the same frequency band and time slot [77]. Thus, FD relay can increase the spectral efficiency compared to its counterpart HD [78]. Therefore, the combination of cooperative communication and NOMA has been considered as a remarkable solution to further enhance the system efficiency of NOMA. Accordingly, in [79], a cooperative transmission scheme, where the users with stronger channel conditions are considered as relays due to their ability in the decoding information of other users in order to assist the users with poor channel conditions, has been proposed to be implemented in NOMA. In [80], by assuming the same scenario in [79], Kim et al. proposed a device-to-device aided cooperative NOMA system, where the direct link is available between the BS and one user, and an upper bound related to sum capacity scaling is derived. In addition, a new power allocation scheme is proposed to maximize the sum capacity. On the other hand, in [81], the authors analyze the performance of NOMA based on user cooperation, in which relaying is realized by one of the users, operating in FD mode to provide high throughput, by applying power allocation.

However, aforementioned user cooperation schemes are more appropriate for short-range communications, such as ultrawideband and Bluetooth. Therefore, in order to further extend the coverage area and to exploit the advantages of cooperation techniques, the concept of cooperative communication, where dedicated relays are used, has also been investigated in NOMA. In this context, in [82], a coordinated transmission protocol where a user communicates with BS directly while the other needs the help of a relay to receive the transmitted information from the BS has been employed in NOMA scheme in order to improve the spectral efficiency, and OP analysis is conducted for frequency-flat block fading channels by using DF relaying, as shown in Figure 8(a). In [83], the same scenario in [82] is considered, and OP and asymptotic expressions are obtained in approximated closed forms for AF relaying networks. Differing from [82] and [83], in [84], the authors proposed a cooperative relaying system, where two symbols transmitted from the BS to the user by the help of a relay were combined at the BS by applying NOMA concept. The exact and asymptotic expressions related to achievable average rate are derived in i.i.d. Rayleigh fading channels and the results demonstrate that cooperative relaying based on NOMA outperforms the conventional one. Also, the authors of [85] analyzed the same transmission scheme in [84] over Rician fading channels. In order to further improve the achievable rate of the system investigated in [84], in [86], authors proposed a novel receiver scheme, where the transmitted symbols from the BS are combined at the destination according to MRC technique and investigated the system performance in terms of ergodic sum rate and OP. Their results demonstrate that the proposed scheme achieves better performance than the one in [84]. In addition, Wan et al. [87] investigated the same system in [86] by using two DF relays and assuming no direct link for cooperation and analyzed the system performance in terms of achievable sum rate. In [88], the authors investigate the performance of NOMA over i.i.d. Rayleigh fading channels by employing a downlink cooperative network in which the BS transmits the superimposed information to the mobile users through a relay and also the direct link is considered. The OP expression of the related user is obtained in closed form, and ergodic sum rate and asymptotic analyses are also maintained as performance metrics. The results show that the NOMA exhibits the same performance in terms of diversity order when compared to OMA by improving spectral efficiency and providing a better user fairness. Furthermore, in [89], performance of NOMA is investigated in relaying networks without the direct link over Nakagami-$m$ fading environments for the network given in Figure 8(b) where all nodes and mobile users are assumed to have a single antenna. While closed-form OP expressions and simple bounds are obtained, ergodic sum rate and asymptotic analyses are also conducted. Under the consideration of imperfect CSI, the authors of [90] analyze the performance of NOMA system investigated in [89] in terms of OP. They provide exact OP and lower bound expressions in closed form and their results show that an error floor comes up due to the imperfect CSI at all SNR region. Similar to the scenario in [89], in [91], performance of NOMA with fixed gain AF relaying is analyzed over Nakagami-$m$ fading channels in case when the direct transmission also exists. For performance criterion, new closed-form expressions related to the exact and asymptotic OPs are obtained. Moreover, a buffer-aided cooperative technique, where the relay transmits and receives the information packets when source-relay and relay-destination links are in outage, respectively, has been taken into account by researchers in order to further enhance the reliability of the relaying systems and increase the system throughput [92]. Accordingly, in [93], the authors proposed a cooperative NOMA system with buffer-aided relaying technique consisting of one source and two users in which the stronger user is used as a buffer-aided relay. Differing from [93], Zhang et al. [94] proposed a buffer-aided NOMA relay network in which a dedicated relay was used to forward the information to two users, and exact OP of the system was obtained in single integral form and lower/upper bounds were derived in closed forms. In [95], for the same system in [94], an adaptive transmission scheme in which the working mode is adaptively chosen in each time slot is proposed to maximize the sum throughput of the considered NOMA system.

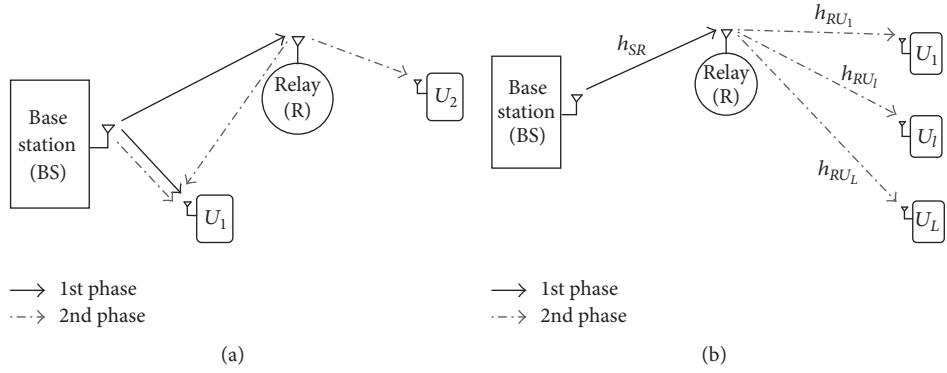

FIGURE 8: System model of cooperative NOMA downlink. (a) Coordinated direct and relay transmission. (b) A cooperative scheme without direct link.

As can be seen from the aforementioned studies, the power allocation issue is vital for the performances of user destinations. In this context, there are several studies that focus on power allocation strategies for cooperative NOMA in the literature [96–99]. Accordingly, in [96], the authors proposed a novel two-stage power allocation scheme for cooperative NOMA with direct link consisting of one source, one relay, and one user destination in order to improve sum rate and OP of the system. In [97], Gau et al. proposed a novel dynamic algorithm that selects the optimal relaying mode and determines the optimal power allocation for cooperative NOMA, where the BS communicates with two users via a couple of dedicated relays. For the proposed approach, new closed-form expressions related to optimal power allocation were derived. In [98], the authors investigated a joint subcarrier pairing and power allocation problem in cooperative NOMA which consists of one BS and two users (one of the users acts as a relay). Theoretical expressions related to joint optimization approach are derived and superiority of the considered algorithms is demonstrated by simulations. In [99], in order to optimize the resource allocation for maximizing the average sum-rate, authors studied the performance of a single-cell NOMA system consisting of multiple source-destination pairs and one OFDM AF relay.

As well known from the literature, diversity techniques and using multiantenna strategies improve system performance significantly. Therefore, in [100], the same authors of [88] consider using multiple antennas at the BS and mobile users and analyze the OP behavior of the network over i.i.d. Rayleigh in case when the direct link does not exist. They apply TAS and MRC techniques at the BS and mobile users, respectively, while the relay has single antenna and show that using multiple antennas improves the system OP performance. Additionally, it is shown that NOMA provides a better OP performance than OMA when the distance between the BS and relay is sufficiently short. In [101], OP performance of the same system investigated in [100] was analyzed for Nakagami-$m$ channels in case that fixed gain AF relay was used. In [102], performance of the same system in [100] was investigated over Nakagami-$m$ fading environments in the presence of imperfect CSI. The system OP was obtained in closed form and tight lower/upper bounds were provided for further insights. In [103], the authors proposed an Alamouti space-time block coding scheme based on two-phase cooperative DF relaying for NOMA and obtained closed-form expressions for both OP and ergodic sum-rate. In [104], the authors analyzed the system performance of nonregenerative massive MIMO NOMA relay network in case that SIC and maximum mean square error SIC techniques were adopted at the receivers. In the system, multiple users and relays are equipped with single antenna while the BS has multiple antennas. For performance metrics, system capacity and sum rate expressions were derived in closed forms and authors demonstrated that the considered system outperforms massive MIMO OMA.

In addition to the aforementioned studies, using multirelays and/or relay selection techniques in cooperative NOMA concept are hot issues since using multiple relays improves the system performance significantly as already known from studies in the literature. Therefore, in [105], the authors proposed a novel NOMA relaying system based on hybrid relaying scheme, where some of relays adopted DF protocol while the others used AF for signal transmission, consisting of two sources and one user destination. For performance comparison with the conventional systems, channel capacity and average system throughput were investigated, and the proposed system was shown to achieve larger sum channel capacity and average system throughput than the conventional systems. Gendia et al. [106] investigated a cooperative NOMA with multiple relays in which all users except the user to whom the information signal would be transmitted were considered as relays. Comparisons with the other equivalent NOMA systems were done in terms of user-average bit error rate, ergodic sum rate, and fairness level by simulations. In [107], OP performance of a NOMA system, where the BS transmits the information signals to two users by using two relays, was analyzed when cooperative and TDMA schemes were applied for transmission. The authors demonstrated that cooperative scheme outperforms TDMA one in terms of OP. Shin et al. [108] proposed a novel multiple-relay-aided uplink NOMA scheme for multicell cellular networks where the BS was equipped with multiantenna and limited by user numbers in each cell. Moreover, the feasibility conditions of the considered system were investigated. Besides multirelaying strategies, relay selection techniques were also investigated. Accordingly, in [109], the authors investigated the impact of two relay selection techniques on the performance of cooperative NOMA scheme without direct link. According to the results, with the relay selection strategies significant performance gain in terms of OP has been achieved in NOMA compared to counterpart OMA. In [110], performance of a cooperative NOMA with the best relay selection technique was analyzed in terms of average rate. The considered relay network consists of one BS, one user, and multiple relays and the direct link is also available. Authors demonstrated that the significant performance gain can be achieved by increasing the number of relays when compared to OMA one. Deng et al. [111] investigated the joint user and relay selection problem in cooperative NOMA relay networks, where multiple source users communicate with two destination users via multiple AF relays. In order to improve the system performance, the authors proposed an optimal relay selection scheme, where the best user-relay pair was selected. In [112], performance of cooperative NOMA with AF relays was analyzed by using partial relay selection technique. In the network, communication between the BS and two users was realized by selected relay, and also direct link between the BS and users was taken into account. While authors provided closed-form OP and sum rate expressions, asymptotic analysis at high SNR region was also conducted. It is shown that the performance can be improved by increasing the number of relays, but the same performance gain is obtained at high SNR region for more than two relays. In addition to above studies, Yang et al. [113] proposed a novel two-stage relay selection scheme for NOMA networks which consists of one source, multiple DF/AF relays, and two users. The considered selection strategy relies on satisfying the QoS of one user in the first stage while maximizing the rate of the other user in the second stage.

Besides that NOMA improves the system spectral efficiency, energy harvesting (EH) technology has also gained much attention because of its ability in increasing energy efficiency. Therefore, simultaneous wireless information and power transfer (SWIPT), which uses radio-frequency signals to enable self-sustainable communication, was proposed by Varshney [114] and regarded as an efficient solution over all emerging EH techniques due to the limitation of environmental energy sources. In this context, many studies combining cooperative NOMA with EH technologies were conducted in the literature [115–123]. In order to exploit the energy and spectral efficiency features of SWIPT and NOMA, Liu et al. [115] studied the application of SWIPT to cooperative NOMA, where users nearby to the BS act as EH relays. In addition, different user selection schemes were proposed in order to determine which nearby user would cooperate with far user, and OP and throughput expressions related to the selection schemes were obtained in closed forms. In [116], a transceiver design problem in cooperative NOMA with SWIPT was studied. In the considered system, the stronger user acting as a relay and BS were equipped with multiple antennas while the other user had only single antenna. Optimal transmitter beamforming and ZF-based transmitter beamforming structures were proposed to maximize the rate of relay node. In [117, 118], the authors analyzed OP performance of NOMA-SWIPT relay networks over i.i.d. Rayleigh and Nakagami-$m$ fading environments, respectively. Differing from the previous works, authors considered that the BS and multiple users were equipped with multiple antennas and communication between the BS and users was established only via an EH relay. They considered that TAS and MRC techniques were employed at the BS and users, respectively, and proved closed-form OP expressions for performance criterion. Similar to [115], in [119], a best-near best-far user selection scheme was proposed for a cellular cooperative NOMA-SWIPT system and OP analysis was conducted to demonstrate the superiority of the proposed scheme. In [120], the authors investigated TAS schemes in MISO-NOMA system based on SWIPT technique, where the BS with multiple antennas communicates with two users with single antenna and the stronger user is also used as an EH relay, in terms of OP and conducted diversity analysis. The impact of power allocation on cooperative NOMA-SWIPT networks was investigated by Yang et al. [121]. For performance comparisons with existing works, OP and high SNR analyses were conducted, and the proposed system was shown to improve the OP performance significantly. In [122], authors analyzed OP performance of a downlink NOMA with EH technique consisting of one BS and two users. While the BS and one of the users which was used as a relay were equipped with multiple antennas, the other user far from the BS had only single antenna. Closed-form OP expressions were derived for AF, DF, and quantize-map-forward relaying protocols over i.i.d. Rayleigh fading channels. Xu et al. [123] investigated joint beamforming and power splitting control problem in NOMA-SWIPT system studied in [120]. In order to maximize the rate of the relay user, power splitting ratio and beamforming vectors were optimized. Moreover, SISO-NOMA system was also studied.

While most of the prior works on the cooperative NOMA systems have focused on the use of HD relaying technique, there are also some studies that consider using FD relaying technique in order to further increase spectral efficiency of NOMA systems. In [124], performance of cooperative SISO-NOMA relaying system consisting of one BS and two users was investigated. The user near BS was considered as an FD relay which employed compress-and-forward protocol for poor user. Authors provided theoretical expressions of achievable rate region based on the noisy network coding. Zhong and Zhang [125] proposed using FD relay instead of HD for the investigated system in [82], where one user can communicate with the BS directly while the other needs a relay cooperation. In order to demonstrate the superiority of using FD relay, authors provided exact OP and ergodic sum capacity expressions. In [126], OP performance of cooperative NOMA system in which the strong user helps the other by acting as an FD-DF relay was analyzed in terms of OP. Moreover, an adaptive multiple access scheme that selects access mode between proposed NOMA, conventional NOMA, and OMA was investigated in order to further enhance the system OP. Differing from [126], authors of [127] investigated optimizing the maximum achievable rate region of cooperative NOMA system in which the BS also operated in FD mode. Therefore, the authors proposed three approaches for maximization problem, such as fixed transmit power, nonfixed transmit power, and transmit power corrupted by error vector magnitude. In [128], a hybrid half/full-duplex relaying scheme was proposed to implement in cooperative NOMA and power allocation problem was investigated in terms of achievable rate. In addition, NOMA with HD and NOMA with FD systems were separately investigated by providing closed-form optimal expressions related to powers. Hybrid NOMA scheme was shown to outperform the other NOMA schemes. The same hybrid NOMA system in [128] was also investigated by Yue et al. [129] in terms of OP, ergodic rate, and energy efficiency. In addition, the authors also investigated the system when the direct link was not available between the BS and poor user. In [130], OP and ergodic sum rate performance of a cooperative NOMA system with FD relaying was investigated in case that the direct link was not available. Theoretical expressions were derived in closed forms. Moreover, in order to maximize the minimum achievable rate, optimization problem for power allocation was also studied.

In the next section, we provide an overview of the cooperative NOMA system which is investigated in [89] to provide an example of cooperative NOMA.

*4.1. Performance Analysis of Cooperative NOMA.* Consider a dual hop relay network based on downlink NOMA as given in Figure 8(b) which consists of one BS ($S$), one AF HD relay ($R$), and $L$ mobile users. In the network, all nodes are equipped with a single antenna, and direct links between the BS and mobile users can not be established due to the poor channel conditions and/or the mobile users are out of the range of BS. We assume that all channel links are subjected to flat Nakagami-$m$ fading. Therefore, channel coefficients of $S$-$R$ and $R$-$U_l$ are denoted by $h_{SR}$ and $h_{RU_l}$ with the corresponding squared means $E[|h_{SR}|^2] = \Omega_{SR}$ and $E[|h_{RU}|^2] = \Omega_{RU}$, respectively, where $l = 1, \ldots, L$. In order to process NOMA concept, without loss of generality, we consider ordering the channel gains of $L$ users as $|h_{RU_1}|^2 \leq |h_{RU_2}|^2 \leq \cdots \leq |h_{RU_L}|^2$. In the first phase, the superimposed signal $s$ given in (1) is transmitted from the BS to the relay and then the received signal at $R$ can be modeled as

$$y_R = h_{SR} \sum_{i=1}^{L} \sqrt{a_i P_s} x_i + n_R, \qquad (39)$$

where $n_R$ is the complex additive Gaussian noise at $R$ and distributed as $CN(0, \sigma_R^2)$.

In the second phase, after the relay applies AF protocol, the received signal at $U_l$ can be written as

$$y_{RU_l} = \sqrt{P_R}Gh_{SR}h_{RU_l}\sum_{i=1}^{L}\sqrt{a_iP_s}x_i + \sqrt{P_R}Gh_{RU_l}n_R \quad (40)$$
$$+ n_{U_l},$$

where $n_{U_l}$ is the complex additive Gaussian noise at $U_l$ and distributed as $CN(0,\sigma_{U_l}^2)$, and $P_R$ is the transmit power at $R$. $G$ denotes the amplifying factor and can be chosen as

$$G = \sqrt{\frac{P_R}{P_s|h_{SR}|^2 + \sigma_R^2}}. \quad (41)$$

In order to provide notational simplicity, we assume that $P_s = P_R = P$, $\sigma_R^2 = \sigma_{U_l}^2 = \sigma^2$. In addition, $\gamma = P/\sigma^2$ denotes the average SNR.

After the SIC process implemented at the receiver of $U_l$, the SINR for the $l$th user can be obtained as [89]

$$\gamma_{RU_l} = \frac{a_l\gamma^2|h_{SR}|^2|h_{RU_l}|^2}{\gamma^2|h_{SR}|^2|h_{RU_l}|^2\Psi_l + \gamma(|h_{SR}|^2 + |h_{RU_l}|^2) + 1}, \quad (42)$$

where $\Psi_l = \sum_{i=l+1}^{L}a_i$. Then, the received SINR by the $L$th user can be simply expressed as [89]

$$\gamma_{RU_L} = \frac{a_L\gamma^2|h_{SR}|^2|h_{RU_L}|^2}{\gamma(|h_{SR}|^2 + |h_{RU_L}|^2) + 1}. \quad (43)$$

Since channel parameters are Nakagami-$m$ distributed, $|\tilde{h}_X|^2$ squared envelope of any unordered link $X$, where $X \in \{SR, RU_l\}$, follows Gamma distribution with CDF

$$F_{|\tilde{h}_X|^2}(x) = \frac{\gamma(m_X, x(m_X/\Omega_X))}{\Gamma(m_X)}$$
$$= 1 - e^{-x(m_X/\Omega_X)}\sum_{n=0}^{m_X-1}\left(\frac{m_X}{\Omega_X}x\right)^n\frac{1}{n!}. \quad (44)$$

In (44), right hand side of the equation is obtained by using the series expansion form of incomplete Gamma function [72, eq. (8.352.6)] and $m_X$ denotes the Nakagami-$m$ parameter belonging to the link $X$.

Furthermore, the PDF and CDF of the ordered squared envelope $|h_X|^2$ can be written by using (44) as [89]

$$f_{|h_X|^2}(x) = Q\sum_{k=0}^{L-l}(-1)^k C_k^{L-l}f_{|\tilde{h}_X|^2}(x)\left[F_{|\tilde{h}_X|^2}(x)\right]^{l+k-1}, \quad (45)$$

$$F_{|h_X|^2}(x) = Q\sum_{k=0}^{L-l}\frac{(-1)^k}{l+k}C_k^{L-l}\left[F_{|\tilde{h}_X|^2}(x)\right]^{l+k}, \quad (46)$$

where $Q = L!/(L-l)!(l-1)!$ and $C_k^K = \binom{K}{k}$ represents the binomial combination.

*4.1.1. Outage Probability of Cooperative NOMA.* By using the approach given in [89], the OP of the $l$th user can be written as

$$P_{\text{out},l} = 1$$
$$- \Pr\left(|h_{RU_l}|^2 > \eta_l^*, |h_{SR}|^2 > \frac{\eta_l^*\left(1+\gamma|h_{RU_l}|^2\right)}{\gamma\left(|h_{RU_l}|^2 - \eta_l^*\right)}\right). \quad (47)$$

The OP expression given in (47) can be mathematically rewritten as

$$P_{\text{out},l} = \underbrace{\int_0^{\eta_l^*}f_{|h_{RU_l}|^2}(x)}_{J_1}$$
$$+ \underbrace{\int_{\eta_l^*}^{\infty}f_{|h_{RU_l}|^2}(x)F_{|h_{SR}|^2}\left(\frac{\eta_l^*(1+\gamma x)}{\gamma(x-\eta_l^*)}\right)dx}_{J_2}. \quad (48)$$

Then, by using (44) and (45), $J_2$ can be calculated as

$$J_2 = 1 - J_1 - Q\sum_{k=0}^{L-l}\sum_{n=0}^{m_{SR}-1}(-1)^k C_k^{L-l}\frac{1}{n!}$$
$$\cdot \int_{\eta_l^*}^{\infty}f_{|\tilde{h}_{RU_l}|^2}(x)\underbrace{\left(F_{|\tilde{h}_{RU_l}|^2}(x)\right)^{l+k-1}}_{\varphi} \quad (49)$$
$$\times e^{-(\eta_l^*(1+\gamma x)m_{SR}/\gamma\Omega_{SR}(x-\eta_l^*))}\left(\frac{\eta_l^*(1+\gamma x)m_{SR}}{\gamma\Omega_{SR}(x-\eta_l^*)}\right)^n dx.$$

In (49), by using binomial expansion [72, eq. (1.111)], $\varphi$ can be obtained in closed form as

$$\varphi = \sum_{t=0}^{l+k-1}\sum_{p=0}^{t(m_{RU}-1)}C_t^{l+k-1}(-1)^t$$
$$\cdot e^{-x(m_{RU}t/\Omega_{RU})}x^p\vartheta_p(t,m_{RU}), \quad (50)$$

where $\vartheta_a(b,g_c)$ denotes multinomial coefficient given in (26).

Furthermore, if we substitute derivative of (44) and (50) into (49) and then by using some algebraic manipulations, $J_2$ can be obtained in closed form. Then, by substituting $J_2$ into (48), we can obtain the OP of $l$th user in closed form as

$$P_{\text{out},l} = 1 - Q\sum_{k,n,t,p,i,q}(-1)^{t+k}C_k^{L-m}C_t^{l+k-1}C_i^n C_q^{p+m_{RU}-1}$$
$$\cdot \frac{\vartheta_p(t,m_{RU})}{n!\Gamma(m_{RU})}\times\left(\frac{m_{RU}}{\Omega_{RU}}\right)^{(2m_{RU}-q+i-1)/2}$$
$$\cdot \frac{\rho^{(i+q+1)/2}}{(t+1)^{(q-i+1)/2}}\left(\frac{m_{SR}}{\Omega_{SR}}\right)^{n-i}\eta_l^{*n-i+p+m_{RU}-1-q} \quad (51)$$
$$\times e^{-(\eta_l^*m_{RU}(t+1)/\Omega_{RU})}e^{-(\eta_l^*m_{SR}/\Omega_{SR})}\times 2$$
$$\times K_{q-i+1}\left(2\sqrt{\frac{\rho m_{RU}(t+1)}{\Omega_{RU}}}\right),$$

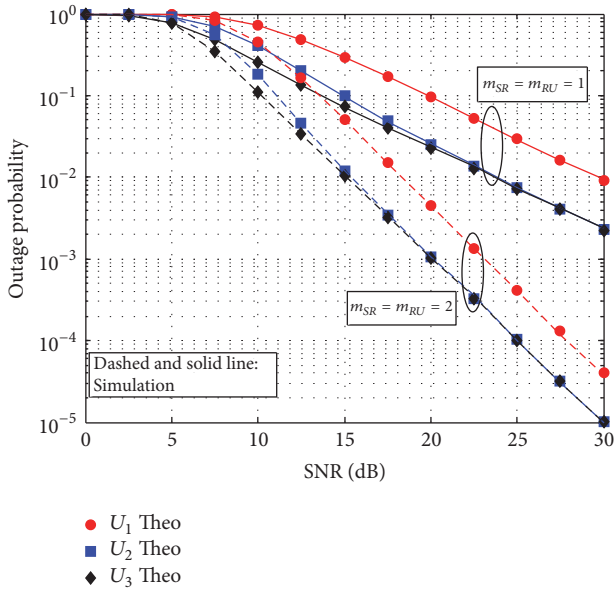

- $U_1$ Theo
- $U_2$ Theo
- $U_3$ Theo

FIGURE 9: Outage probability of NOMA versus SNR in case $d_{SR} = 0.5$ and different Nakagami-$m$ parameters.

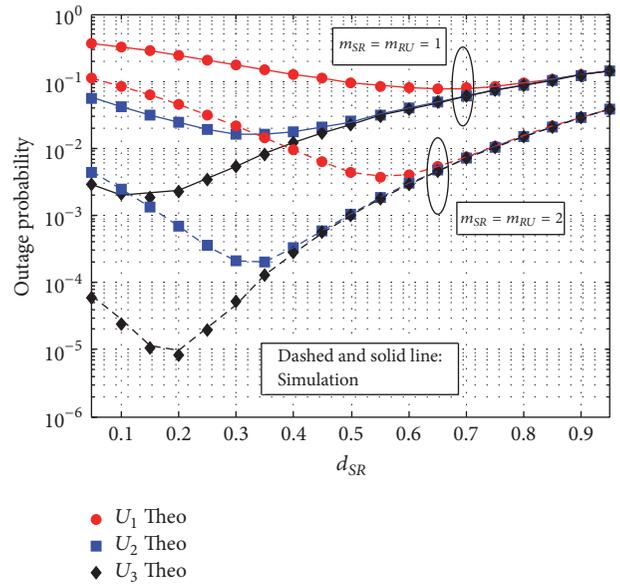

- $U_1$ Theo
- $U_2$ Theo
- $U_3$ Theo

FIGURE 10: Outage probability of NOMA versus $d_{SR}$ in case $\gamma = 20$ dB and different Nakagami-$m$ parameters.

where the binomial expansion [72, eq. (1.111)] and the integral representation in [72, eq. (3.471.9)] are used for the derivation. In (51), $\rho = \eta_l^* m_{SR}(1 + \gamma \eta_l^*)/\gamma \Omega_{SR}$ and $\sum_{k,n,t,p,i,q} \equiv \sum_{k=0}^{L-l} \sum_{n=0}^{m_{SR}-1} \sum_{t=0}^{l+k-1} \sum_{p=0}^{t(m_{RU}-1)} \sum_{i=0}^{n} \sum_{q=0}^{p+m_{RU}-1} (\cdot)$ notations are used to provide a short hand representation. $K_\nu(\cdot)$ denotes the $\nu$th order modified Bessel function of second kind [72, eq. (8.407.1)]. The OP expression in (51) is in a simpler form when compared to equivalent representations in the literature.

*4.1.2. Numerical Results of Cooperative NOMA.* In this section, we provide numerical examples of the provided theoretical results obtained for the OP of NOMA and validate them by Monte Carlo simulations. We assume that the distances between the BS and the mobile users are normalized to one, so that $\Omega_{SR} = d_{SR}^{-\kappa}$ and $\Omega_{RU} = (1 - d_{SR})^{-\kappa}$, where $\kappa = 3$ is the path loss exponent. In all figures, $L = 3$ users and $a_1 = 1/2$, $a_2 = 1/3$, $a_3 = 1/6$, $\gamma_{th_1} = 0.9$, $\gamma_{th_2} = 1.5$, $\gamma_{th_3} = 2$ parameters have been used.

In Figure 9, we present the OP performance of NOMA versus SNR. As can be seen from the figure, theoretical results are well matched with simulations. In addition, OP performances of the second and third users are better than that of the first user and also the same at high SNR region. Moreover, as the channel parameters increase, the OPs of all users increase.

Figure 10 plots the OP performance of NOMA versus the normalized distance between the BS and the relay. As seen from the figure, while the optimal relay location of the user with the strongest channel condition is near the BS, the other users' optimal relay locations are far from the BS since the user with worse channel has higher power allocation coefficient.

## 5. Practical Implementation Aspects

In the literature, power allocation and user clustering are generally considered as the main problems in NOMA systems, and several strategies are proposed to provide efficient solutions to these issues. As also considered in [131–133], these problems are formulated as an optimization problem and the corresponding solution procedures are also proposed. Besides these, studies, such as [54, 134, 135], propose approaches that are suitable to real-time applications. Imperfect CSI is assumed in the corresponding system models. However, real-time implementation challenges are not considered in most of the studies and the associated implementation design, which may provide effective solutions to these challenges, is not mentioned. In this section, these challenges are highlighted and important design components are explained. In the following subsection, some studies that include real-time implementation of NOMA are mentioned and challenges of such real-time implementations will be detailed.

*5.1. Related Works.* The number of studies that target real-time implementation of NOMA is very limited. To the best of the authors' knowledge, beyond three main studies, such content is not included in any other study at the time of preparation of this paper. In [136], single user- (SU-) MIMO is integrated to downlink and uplink NOMA, and extensive computer simulations provide detailed rate evaluation between OMA and NOMA methods. Moreover, a comprehensive testbed is created to experiment downlink NOMA with SU-MIMO setup under real-time impairments. Turbo encoding is also utilized in the implementation and a SIC decoding structure, which also includes turbo decoding and MIMO detection, is proposed. Due to usage of a wider bandwidth, NOMA provides data rate improvement of 61% in this experiment scenario. Reference [137] targets improper power allocation issue, which is seen as a performance limiting factor in conventional NOMA models. By exploiting the physical-layer network coding (PNC) in NOMA, the authors propose network-coded multiple access (NCMA). Adaptation of PNC provides an additional transmission dimension, and the received signals via two different dimensions increase the throughput significantly when compared to the conventional NOMA systems. It is validated by experimental results that the proposed NCMA variations provide noticeable performance improvements under the power-balanced or near power-balanced scenarios. As the final study, in [138], software defined radio (SDR) implementation of downlink NOMA is realized to evaluate the performance differences between NOMA and OMA techniques. Moreover, protocol stack of LTE is modified to propose a suitable protocol stack for NOMA. Besides these multilayer modifications, detailed experiments are also carried out. Measurement results demonstrate the performance advantages of NOMA over OMA.

Since superposition coding and NOMA are very similar in context, studies on superposition coding also contain the same valuable outcomes. In [139], advantages of superposition coding over time division multiplexing approach in terms of improving the quality of the poor links are validated via an SDR platform. Accordingly, the packet error rate is measured and need of a joint code optimization is shown. Moreover, an improved packet error rate performance that is obtained with superposition coding, when compared to the results of time division multiplexing utilization, is demonstrated. Similarly in [140], the authors propose a scheduler based on superposition coding and it is demonstrated that superposition coding based resource allocation can provide a data rate improvement up to 25% when compared to the orthogonal access techniques.

These studies provide significant insights about real-time implementation aspects of NOMA. However, several practical challenges are not yet considered in available works.

*5.2. Implementation Challenges.* Practical implementation challenges of NOMA are considered in some surveys. In [141], the authors focus on multicell NOMA and the related design issues in the environment in the presence of a strong intercell interference (ICI). Since future wireless networks are expected to be densely deployed, NOMA technique is considered to be a candidate technique. ICI should be considered due to the potential effects of interference between adjacent BSs. Theoretical details of single-cell and multicell NOMA solutions are detailed and the capacity analysis is provided. Moreover, some major implementation issues are highlighted. Hardware complexity and error propagation issues of SIC implementation are detailed. Then, the importance of CSI is highlighted and the damaging effects of imperfect CSI on the performance of NOMA are explained. Multiuser power allocation and clustering are also emphasized. To limit ICI between adjacent cells, the authors propose that users should be clustered properly and power allocation mechanism should be operated efficiently. Integration of fractional frequency reuse with NOMA is also considered as a major challenge and such integration should be allocated properly to obtain significant gains. Lastly, security is highlighted as another challenge, and the implementation of physical layer security techniques is seen as a difficult task. As demonstrated with computer simulations targeting to demonstrate the performance limitation of interference, proper ICI cancellation is very significant to obtain a robust performance in multicell NOMA systems.

In [142], challenges of downlink and uplink NOMA implementations and their implementation differences are explained. As the first challenge, implementation complexity is highlighted, where it is pointed out that downlink NOMA brings more complexity because of the utilization of iterative detection procedures multiple times at multiple receive nodes, when compared to the central receiver node, as applicable in uplink NOMA systems. Secondly, intracell/intracluster interference is stated as a crucial issue for both systems due to interference effects between users. As the third challenge, SIC receivers which are implemented differently in downlink and uplink cases are considered. Lastly, ICI is elaborated. It is shown that ICI is more effective in uplink case and could limit performance significantly. However, it is not that effective in downlink case and the observed performance degradation is comparable to that of observed in OMA systems. Moreover, some critical points are listed. Firstly, propagation errors in SIC receivers are mentioned as an important performance limiting factor and interference cancellation schemes are considered necessary to improve these effects. Secondly, multicell NOMA is highlighted, where obtaining the same single-cell NOMA gains over OMA in multicell scenarios becomes challenging. User grouping/scheduling, power allocation, and ICI mitigation are also considered as crucial items to obtain an improved performance. Besides these implementation issues, integration of NOMA-based wireless backhauling to small cells and cooperative schemes are highlighted as necessary precautions to increase NOMA's applicability in real-time.

In [143], implementation issues of NOMA are discussed and listed. Decoding complexity, error propagation, and errors that faced power balanced scenarios are also mentioned. As less considered issues, quantization errors that lead to degradation of weak signals, power allocation complexity due to difficulty of optimization of proper power levels to all users, residual timing offset that leads to synchronization loss, and error increment are highlighted. Furthermore, signaling and processing overhead due to learning procedure of CSI are also listed as a critical inefficiency source.

Some of the main problems that are mentioned in these studies and other issues that are not yet discussed in the literature will be listed and detailed below.

*(1) Hardware Complexity.* When compared to OMA, NOMA causes increased complexity on the hardware side due to SIC implementation. To obtain the users' symbols that transmit or receive with lower power symbols, high power symbols are required to be estimated first with the SIC detector. If the number of users especially is high or fast signal transmission is required, the SIC procedure that is used multiple times, in addition to the detection delay, could cause important limitations for battery-limited devices. Since longer battery life is desired in consumer electronics, implementation of NOMA, particularly in dense networks, could be inefficient. This issue may limit usage of NOMA. Effective user clustering and power allocation are crucial to alleviate this problem.

*(2) Error Propagation in SIC Implementation.* According to the main principle of NOMA, on the receiver side, the user with better channel conditions is estimated first via SIC detection. Therefore, the success of the reception of main signal depends on successful estimation of the high power signals. Since channel and hardware impairments are effective in the reception process, SIC detection can be negatively affected. It is not straightforward for NOMA systems to ideally estimate channel, due to the presence of carrier frequency offset (CFO), timing offset (TO), and other hardware related impairments. Thus, erroneous detection and error propagation are probable in the SIC detection process. To overcome this and to improve the transmission quality, more robust solutions are necessary. Rather than changing the main detector components, improving the estimation quality of mentioned impairments is a more effective approach to obtain a practical performance gain.

*(3) Optimal Pilot Allocation.* Since multiple signals are transmitted in an overlapped fashion, interference emerges and error performance starts to degrade in NOMA, when compared to OMA systems. It is a clear fact that perfect or near-perfect CSI is a must to obtain a good performance. Pilot positions and the number of allocated pilots are important design considerations in NOMA implementation. These are critical even in OMA systems due to uncertain channel characteristics in wireless communication environments. However, due to the inherent interference, optimal pilot allocation is more critical for NOMA systems and careful design is required. Therefore, channel characteristics should be tracked efficiently and accurately to allocate sufficient number of pilots at proper positions, which could result in good error performance in NOMA systems.

*(4) Instantaneous CSI Requirement.* Besides pilot allocation issues in NOMA implementations, another basic CSI estimation issue exists in this process. Allocation of a previously allocated frequency band to a secondary user brings a serious problem; CSI for the transmission of this user should be estimated with orthogonal transmissions. This inevitably blocks the transmission of main user and results in an unfavorable situation. It is not clear whether this issue can be tolerated or not in real-time. Moreover, in dense networks, instantaneous band allocation may be required and, in these cases, this issue may become more critical. Effective and practical solution to this problem is very important for the future of NOMA systems. As a road map suggestion, pilot contamination problem in massive MIMO systems may be considered and corresponding solutions like [144] may be applied to NOMA systems. However, differences between the logics of these techniques should also be taken into account.

*(5) Carrier Frequency Offset and Timing Offset Estimation.* Due to the nature of wireless devices, CFO and TO emerge frequently during communication. Low-quality clocks especially that are included in such devices cause significant CFO and TO, thus, leading to a significantly degraded transmission quality. Usage of multicarrier waveforms like OFDM renders robust CFO and TO estimation and provides the necessary correction. In the point-to-point OMA transmissions, joint estimation of CFO and TO is quite straightforward due to distinguishability of received signals. Even in these cases, these impairments could cause serious performance degradation. However, this is not valid for NOMA because of the reception of signals in an overlapped fashion. This issue has not yet been considered in the literature. Effective solutions and practical approaches are required to guarantee a good transmission quality in NOMA. Highly accurate synchronization support to devices can overcome such disturbances; however, lower cost expectations prevent such a solution. Therefore, particularly, in uplink transmissions, distinguishability of overlapped signals should be achieved.

*5.3. Lessons Learned.* In order to capture the full set of advantages of NOMA in real-time that are validated in the theoretical studies, possible major challenges should be investigated and a comprehensive implementation strategy that overcomes these challenges should be determined. There are few studies in the literature that list these challenges, but there are some challenges that have not yet been considered. From this perspective, in this section, previously mentioned challenges are evaluated and important ones are given with other undetected major challenges. These also provide topics that deserve attention from the researchers who target improving NOMA's applicability.

## 6. Conclusion

NOMA schemes are proposed to improve the efficient usage of limited network sources. OMA based approaches that use time, frequency, or code domain in an orthogonal manner cannot effectively utilize radio resources, limiting the number of users that can be served simultaneously. In order to overcome such drawbacks and to increase the multiple access efficiency, NOMA technique has been recently proposed. Accordingly, users are separated in the power domain. Such a power-domain based multiple access scheme provides effective throughput improvements, depending on the channel conditions.

In OMA, differences between channels and conditions of users cannot be effectively exploited. It is quite possible for a user to be assigned with a large frequency band while experiencing deteriorating channel conditions. Such user cases limit the effectiveness of OMA based approaches. However, according to the NOMA principle, other users who may be experiencing better channel conditions can use these bands and increase their throughput. Moreover, corresponding users who are the primary users of these bands continue to use these bands. In such deployments, power level of users is selected in a way to target a certain maximum error rate. Furthermore, the performance of NOMA can be significantly improved using MIMO and cooperative communication techniques.

In this paper, we provide a unified model system model for NOMA, including MIMO and cooperative communication scenarios. Implementation aspects and related open issues are detailed. A comprehensive literature survey is also given to provide an overview of the state-of-the-art.

## Conflicts of Interest

The authors declare that they have no conflicts of interest.

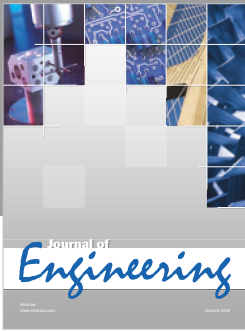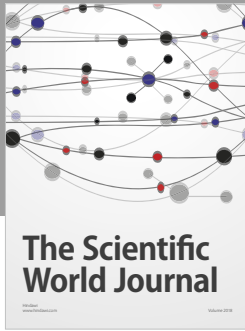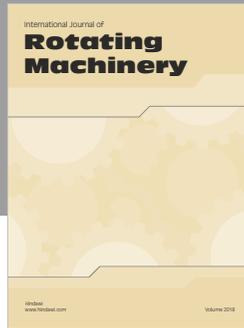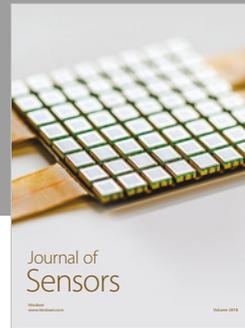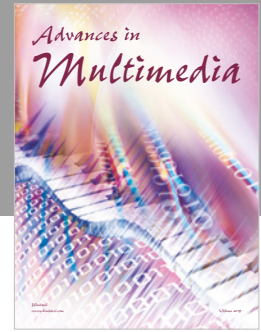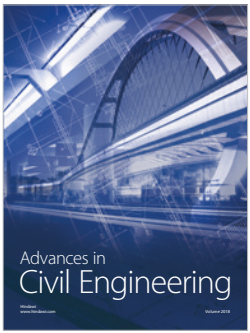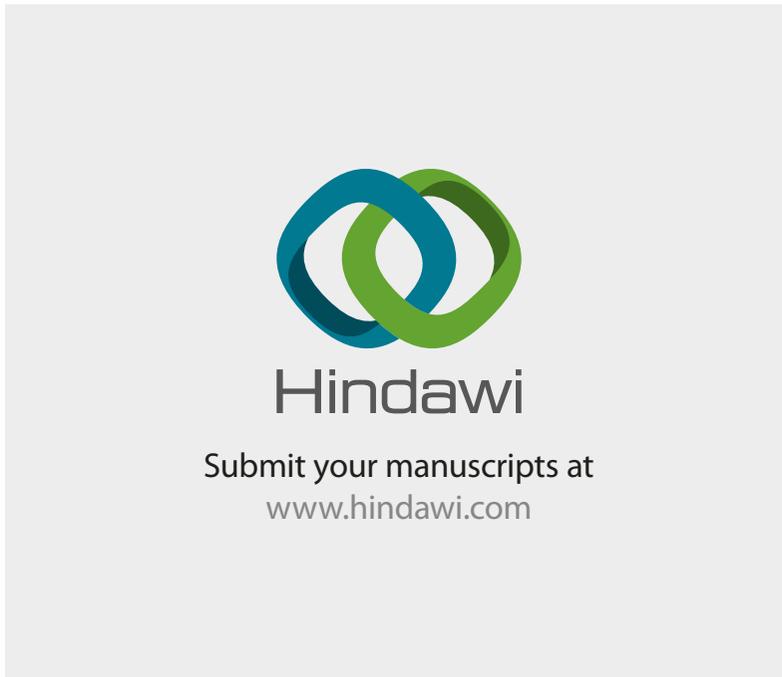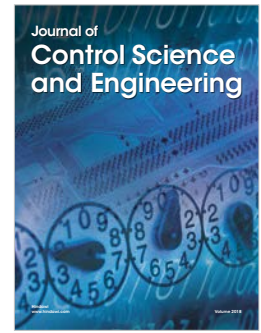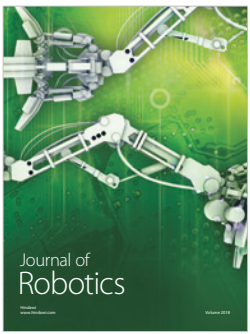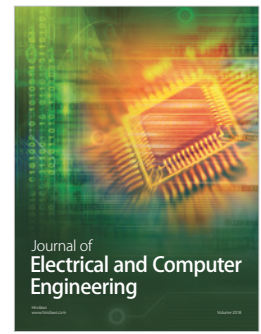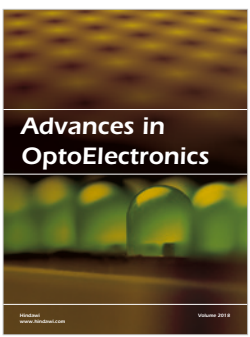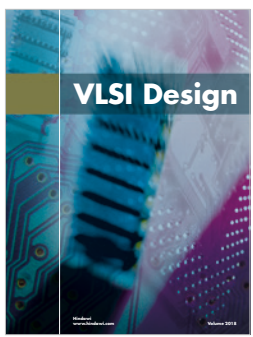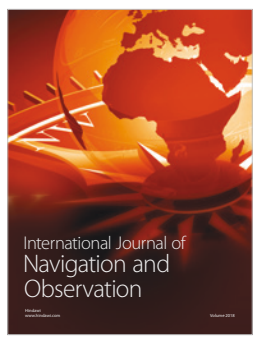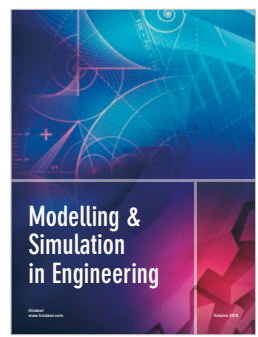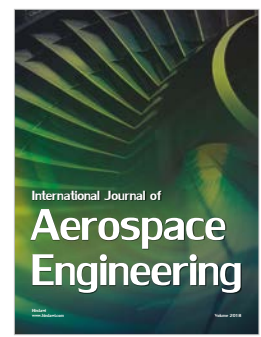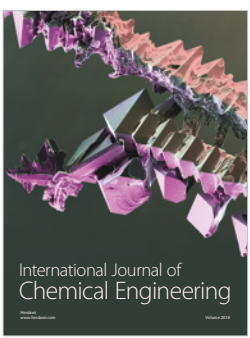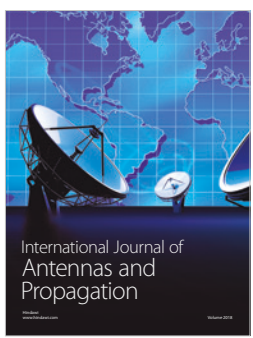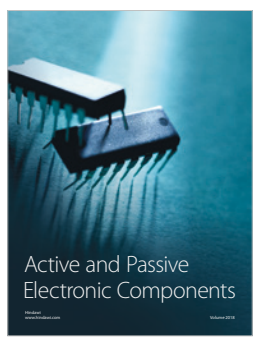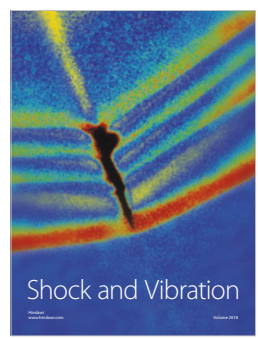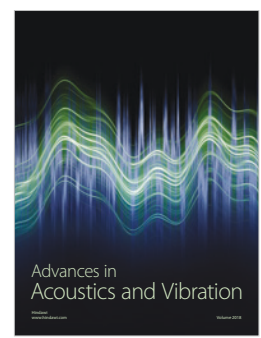